\documentclass[apj,iop]{emulateapj2}

\newcommand{\um}{\ensuremath{{\mu}m}}
\newcommand{\uJy}{\ensuremath{{\mu}Jy}}

\newcommand{\swire}{{\tt SWIRE}}
\newcommand{\cosmos}{{\tt COSMOS}}

\newcommand{\goods}{{\tt GOODS}}
\newcommand{\All}{{\tt All}} 
\newcommand{\COSMOSA}{{\tt COSMOS$_{\tt Shallow}$}}
\newcommand{\ESONE}{{\tt ES1}}
\newcommand{\XMM}{{\tt XMM}}

\newcommand{\Stern}{{Stern et al. }}
\newcommand{\Donley}{{Donley et al. }}

\newcommand{\spitzer}{\textit{Spitzer}}
\newcommand{\chandra}{\textit{Chandra}}
\newcommand{\xmm}{\textit{XMM-Newton}}
\newcommand{\ergscm}{$\mathrm{erg\;s^{-1} cm^{-2}}$}
\newcommand{\cmsq}{\ensuremath{\mathrm{cm}^{-2}}}
\newcommand{\ergs}{\ensuremath{\mathrm{erg\;s^{-1}}}}

\newcommand{\degsq}{\ensuremath{\mathrm{deg}^2}}
\newcommand{\msun}{\mathcal{M}_{\sun}}

\newcommand{\fx}[2]{$f_\mathrm{X} {#1} {10}^{#2}$~\ergscm}
\newcommand{\fhard}[3]{$f_{\mathrm{{#1}keV}}\sim{#2}\times{10^{#3}}$~\ergscm}
\newcommand{\LX}{\ensuremath{L_\mathrm{X}}}
\newcommand{\Lx}[2]{$\LX\;{#1}\;{10}^{#2}$\;\ergs}
\newcommand{\fIR}{\ensuremath{f_{5.8\um}}}
\newcommand{\fir}[2]{$\fIR\;{#1}\;{#2}$~\uJy}
\newcommand{\LIR}{\ensuremath{L_{3.6\um}}}
\newcommand{\Lir}[2]{$\LIR\;{#1}\;{10}^{#2}$~\ergs}

\newcommand{\Lratio}{\LIR/\LX}

\newcommand{\mstar}{\ensuremath{\mathcal{M}_*}}
\newcommand{\mass}[2]{\mstar$\;{#1}\;{10}^{#2}~\msun$}
\newcommand{\mrange}[2]{$ {10}^{#1} < \mstar/~\msun < {10}^{#2}$}

\newcommand{\Nh}[2]{$N_\mathrm{H}\;{#1}\;10^{#2}$~\cmsq}
\newcommand{\LogNLogS}{log\,\textit{N}\ -- log\,\textit{S}}

\newcommand{\specific}{\ensuremath{\lambda}}
\newcommand{\Specific}[2]{\ensuremath{\specific\ {#1}\ {10}^{#2}}}
\newcommand{\SpecificP}[2]{\ensuremath{\specific\ {#1}\ {#2}\%}}

\usepackage{natbib}
\usepackage[usenames]{color}
\definecolor{red}{rgb}{0.75,0,0}
\definecolor{green}{rgb}{0,0.5,0}
\definecolor{blue}{rgb}{0,0,0.75}
\usepackage[ colorlinks=true, 
             urlcolor=blue,
             filecolor=blue,
             linkcolor=red,
             citecolor=blue,
             pdftitle={PRIMUS: IR-AGN and X-ray AGN},
             pdfauthor={Alexander Mendez, et al.},
             bookmarks,
             bookmarksopen=true]{hyperref}

\usepackage{microtype} 
\usepackage{amsmath}   
\usepackage{makecell} 
\usepackage{rotating} 

\newcommand{\nPrimusRedshifts}{1,541}
\newcommand{\nPrimusRedshift}{31,998}

\newcommand{\nCosmosIracCsnr}{3, 3, 16, and 50~\uJy}

\newcommand{\nCosmosXrayIR}{2,769}
\newcommand{\nCosmosXrayPrimus}{2,075}
\newcommand{\nCosmosXrayRedshift}{383}
\newcommand{\nElaisXrayArea}{2.16~\degsq}

\newcommand{\nElaisXrayPrimus}{371}
\newcommand{\nElaisXrayIR}{395}
\newcommand{\nElaisXrayRedshift}{73}
\newcommand{\nXmmXrayArea}{2.16~\degsq}
\newcommand{\nXmmXrayIRArea}{2.16~\degsq}
\newcommand{\nXmmXrayPrimus}{1840}
\newcommand{\nXmmXrayIR}{4466}
\newcommand{\nXmmXrayRedshift}{264}

\newcommand{\nSimpleFluxCorr}{2.02, 1.96, 1.69, and 1.56}

\newcommand{\nCdfsXrayEcdfsSimpleArea}{0.068~\degsq}
\newcommand{\nCdfsXrayEcdfsSwireArea}{0.12~\degsq}
\newcommand{\nCdfsXrayCdfsSimpleArea}{0.25~\degsq}
\newcommand{\nCdfsXrayCdfsSwireArea}{0.31~\degsq}
\newcommand{\nCdfsXrayCdfsGoodsArea}{0.058~\degsq}
\newcommand{\nSternIRObscuredFraction}{3\%$\pm$1\%}
\newcommand{\nDonleyIRObscuredFraction}{6\%$\pm$3\%}
\newcommand{\nExcessCut}{-2.0}
\newcommand{\nSternBrightLogXLum}{79\%$\pm$4\%}
\newcommand{\nDonleyBrightLogXLum}{46\%$\pm$6\%}

\newcommand{\nXrayBrightHRObscuredFraction}{28\%$\pm$7\%}
\newcommand{\nSternBrightHRObscuredFraction}{18\%$\pm$10\%}
\newcommand{\nDonleyBrightHRObscuredFraction}{20\%$\pm$17\%}
\newcommand{\nXrayFaintHRObscuredFraction}{28\%$\pm$5\%}
\newcommand{\nSternFaintHRObscuredFraction}{36\%$\pm$15\%}
\newcommand{\nDonleyFaintHRObscuredFraction}{67\%$\pm$27\%}

\newcommand{\nSternMassiveFraction}{52\%$\pm$2\%}
\newcommand{\nDonleyMassiveFraction}{69\%$\pm$4\%}
\newcommand{\nXrayMassiveFraction}{78\%$\pm$2\%}

\newcommand{\nSternMediumFraction}{21\%$\pm$2\%}
\newcommand{\nDonleyMediumFraction}{11\%$\pm$3\%}
\newcommand{\nXrayMediumFraction}{7\%$\pm$1\%}

\newcommand{\nSternObscuredFraction}{48\%$\pm$3\%}
\newcommand{\nSternAllXrayMissingFraction}{27\%$\pm$4\%}
\newcommand{\nSternLowXrayMissingFraction}{76\%$\pm$10\%}
\newcommand{\nSternHighXrayMissingFraction}{86\%$\pm$13\%}

\newcommand{\nDonleyBroadLineFraction}{32\%$\pm$3\%}
\newcommand{\nSternBroadLineFraction}{16\%$\pm$2\%}
\newcommand{\nXrayBroadLineFraction}{12\%$\pm$1\%}

\newcommand{\nTotalStern}{9005}
\newcommand{\nTotalDonley}{3569}
\newcommand{\nTotalXray}{4886}

\newcommand{\nSwireSternTotal}{6687}
\newcommand{\nCosmosSternTotal}{10,343}
\newcommand{\nGoodsSternTotal}{1137}

\newcommand{\nSwireDonleyTotal}{4151}
\newcommand{\nCosmosDonleyTotal}{2657}
\newcommand{\nGoodsDonleyTotal}{212}

\begin{document}
\title{PRIMUS: Infrared and X-ray AGN Selection Techniques at $0.2<z<1.2$}
\shorttitle{PRIMUS: IR-AGN and X-ray AGN}
\shortauthors{Mendez et al.}

\author{Alexander J. Mendez\altaffilmark{1},
        Alison L. Coil\altaffilmark{1,9},
        James Aird\altaffilmark{1,2},
        Aleksandar M. Diamond-Stanic\altaffilmark{1,10},
        John Moustakas\altaffilmark{3},
        Michael R. Blanton\altaffilmark{4},
        Richard J. Cool\altaffilmark{5,11},
        Daniel J. Eisenstein\altaffilmark{6},
        Kenneth C. Wong\altaffilmark{7},
        Guangtun Zhu\altaffilmark{8}}
\altaffiltext{1}{Center for Astrophysics and Space Sciences, Department of Physics, University of California, 9500 Gilman Dr., La Jolla, San Diego, CA 92093, USA}
\altaffiltext{2}{Department of Physics, Durham University, Durham DH1 3LE, UK}
\altaffiltext{3}{Department of Physics and Astronomy, Siena College, 515 Loudon Road, Loudonville, NY 12211, USA}
\altaffiltext{4}{Center for Cosmology and Particle Physics, Department of Physics, New York University, 4 Washington Place, New York, NY 10003, USA}
\altaffiltext{5}{The Observatories of the Carnegie Institution for Science, 813 Santa Barbara Street, Pasadena, CA 91101, USA}
\altaffiltext{6}{Harvard College Observatory, 60 Garden St., Cambridge, MA 02138, USA}
\altaffiltext{7}{Steward Observatory, The University of Arizona, 933 N. Cherry Ave., Tucson, AZ 85721, USA}
\altaffiltext{8}{Department of Physics and Astronomy, Johns Hopkins University, 3400 N. Charles Street, Baltimore, MD 21218, USA}
\altaffiltext{9}{Alfred P. Sloan Foundation Fellow}
\altaffiltext{10}{Center for Galaxy Evolution Fellow}
\altaffiltext{11}{Hubble Fellow, Princeton-Carnegie Fellow}

\begin{abstract}
We present a study of \spitzer/IRAC and X-ray active galactic nucleus (AGN) selection techniques in order to quantify the overlap, uniqueness, contamination, and completeness of each.
We investigate how the overlap and possible contamination of the samples depends on the depth of both the IR and X-ray data.
We use \spitzer/IRAC imaging, \chandra\ and \xmm\ X-ray imaging, and spectroscopic redshifts from the PRism MUlti-object Survey (PRIMUS) to construct galaxy and AGN samples at $0.2 < z < 1.2$ over 8~\degsq. 
We construct samples over a wide range of IRAC flux limits (SWIRE to GOODS depth) and X-ray flux limits (10~ks to 2~Ms). 
We compare IR-AGN samples defined using both the IRAC color selection of \citeauthor{Stern05} and \citeauthor{Donley12} with X-ray detected AGN samples.
For roughly similar depth IR and X-ray surveys, we find that $\sim$75\% of IR-selected AGNs are also identified as X-ray AGNs.
This fraction increases to $\sim$90\% when comparing against the deepest X-ray data, indicating that at most $\sim$10\% of IR-selected AGNs may be heavily obscured.
The IR-AGN selection proposed by \Stern suffers from contamination by star-forming galaxies at various redshifts when using deeper IR data, though the selection technique works well for shallow IR data.
While similar overall, the IR-AGN samples preferentially contain more luminous AGNs, while the X-ray AGN samples identify a wider range of AGN accretion rates including low specific accretion rate AGNs, where the host galaxy light dominates at IR wavelengths.
The host galaxy populations of the IR and X-ray AGN samples have similar restframe colors and stellar masses; both selections identify AGNs in blue, star-forming and red, quiescent galaxies.
\end{abstract}

\section{Introduction}\label{sec:intro}
Understanding the nature and role of active galactic nuclei (AGNs) is crucial for understanding both the accretion history of the universe as well as galaxy evolution.
There is mounting observational evidence that there is a connection between black hole growth and galaxy growth.
This is shown both by the tight correlation between black hole mass and galaxy bulge mass \citep[e.g.,][]{Magorrian98, Ferrarese00, Gebhardt00}, as well as the similar evolutionary history of star formation and AGN activity through cosmic time \citep[e.g.,][]{Boyle98, Silverman08,Aird10}.
In order to understand and characterize AGNs, one must be able to identify a complete AGN sample, with full knowledge of any underlying biases or contamination in the sample.
One can then better determine which processes (e.g. secular evolution, mergers, and environment) are the dominate fueling mechanisms in the growth and evolution of AGNs.

Deep X-ray surveys provide a reliable means of selecting AGNs, in that they introduce few false positives, as the AGN light generally outshines light from even highly active star-forming galaxies at X-ray wavelengths.
However, high-column density gas (\Nh{>}{23}) absorbs X-rays, such that X-ray surveys may fail to identify the most heavily absorbed AGNs.
While the exact fraction of obscured AGNs (\Nh{>}{22}) is not yet fully known, obscured AGNs likely represent a large fraction of the total AGN population at all luminosities.
For example, \citet{Treister04} predict that between $\sim$ 25\% to $\sim$ 50\% of AGNs are obscured, even at high luminosities (\Lx{>}{44}); \citet[see also ][]{Treister09, Treister09b, Ballantyne11}, while \citet{Gilli07} use the cosmic X-ray background to predict that Compton-thick AGNs are four times as numerous as unobscured AGNs at low luminosity (\Lx{<}{43.5}).
More recently, \citet{Akylas12} predict that the number of Compton-thick AGNs may be 10 times as numerous as unobscured AGNs.
Only with extremely deep X-ray data ($\gtrsim$1 Ms) does it become possible to detect and characterize the heavily obscured, moderate luminosity AGNs \citep[e.g.][]{Georgantopoulos09, Brightman12} that would otherwise be missed in shallower X-ray surveys. 
However, even the deepest X-ray surveys may still fail to identify low luminosity AGNs with moderate to heavy obscuration.
Additionally, current deep X-ray surveys cover at most $\sim$1~\degsq\ of sky, which limits the size of the resulting AGN samples.
Ideally, both wide and deep X-ray surveys are required to identify statistically large, relatively complete AGN samples.

The use of mid-infrared (MIR) emission to identify AGNs began in the 1970s with the advent of sensitive IR detectors \citep{Low68} and continued with the \textit{Infrared Astronomical Satellite} and the Infrared Array Camera \citep[IRAC; ][]{Fazio04} on board \spitzer. 
High-energy radiation from the AGN is reprocessed by dust near the AGN and re-radiated at MIR wavelengths. 
Luminous AGNs display a red MIR power-law spectral energy distribution (SED), which is dominated by thermal emission from hot dust \citep{Neugebauer79, Elvis94, Rieke81}.
Emission at MIR wavelengths can also potentially be used to detect Compton-thick AGNs that may be missed by deep X-ray surveys \citep[e.g.][]{Ivison04, Lacy04, Stern05, Alonso-Herrero06, Polletta06}.
It has also been found that objects with similar MIR luminosities can have vastly different radio, optical, UV and soft X-ray luminosities \citep{Mushotzky04}, such that heavily obscured AGNs can be identified in the MIR as it is relatively insensitive to obscuration.
Therefore MIR AGN identification is a potentially powerful tool that is sensitive to both obscured and unobscured AGNs, without requiring time-consuming deep X-ray data.
Building on the above findings, a variety of selection techniques have been developed to identify large samples of AGNs from MIR imaging data.
Within the MIR waveband there are many techniques to select infrared-AGNs (IR-AGNs), each taking advantage of the unique colors of AGNs.
\citet{Stern05} and \citet{Lacy04} propose IRAC color-color cuts that were designed using shallow IRAC surveys and which effectively select luminous AGNs at low redshift.
These selections identify the AGN population and separate them from the much larger galaxy population based on their characteristic MIR properties.
Recently, a number of new MIR selection techniques have been suggested that use data from the \textit{Wide field Infrared Survey Explorer} \citep[\textit{WISE}; ][]{Wright10} survey, including those of \citet{Messias12}, who use the 4.5~\um and 8.0~\um\ bands, and \citet{Mateos12}, who use the 2.4~\um, 4.6~\um, and 12~\um\ bands.
The \textit{WISE} bands used in these results are not very different from the \spitzer\ IRAC bands and add techniques to fully probe the AGN population space.
Alternatively, the IR SED can be fit using multiple bands and the associated photometric errors to estimate the probability that the source has a featureless power-law continuum \citep[e.g.][]{Alonso-Herrero06, Polletta06, Donley07}.
Such techniques identify AGNs within similar regions in color--color space but are generally more reliable and lead to smaller samples. 

Recent work using deeper IRAC and X-ray surveys \citep[e.g.][]{Barmby06, Donley07, Cardamone08, Park10, Eckart10} has begun to investigate these MIR AGN selections beyond the shallow surveys for which they were designed.
\citet{Barmby06} use deep IR and X-ray data in the Extended Groth Strip (EGS) to find that X-ray AGN have a wide range of MIR colors, suggesting that there is no single method that will be able to identify a complete AGN sample. 
Likewise, \citet{Park10} find that a majority (78\%) of the X-ray AGN sources are not detected by the power-law AGN selection method.
Together these studies suggest that there is a population of AGNs missed by IR-AGN selection techniques even at shallow X-ray and IR survey depths.
Recently, \citet{Donley12} revisit the \Stern and Lacy IRAC color-color selections, which were defined with relatively shallow surveys, and the power-law selection, which depends on both the photometry and (generally underestimated) photometric errors, and define a \textit{new} IRAC color-color selection criteria, which they claim is more reliable than the previous IR selection criteria.
Additionally, using an X-ray stacking analysis \citet{Georgantopoulos08} find a soft mean X-ray spectrum in the Chandra Deep Field North (CDFN) suggesting contamination of the Stern wedge by normal galaxies.
While each of these IR AGN criteria select some fraction of the total underlying AGN population, \citet{Barmby06} suggest that no proposed IR AGN color selection will identify \textit{all} AGNs, given the wide range of spectral shapes exhibited by X-ray sources in the EGS.

With some possible issues, using the MIR to select AGNs is still an effective way to identify heavily obscured/Compton-thick AGNs that are an important population missed by X-ray surveys. 
In order to build more complete AGN samples it has become common to use multiple waveband selection techniques in the IR through X-ray.
For example, \citet{Hickox07}, and \citet{Assef10} combine different waveband selection techniques to probe the properties of the underlying AGN population.
Combining both X-ray AGN samples and IR-AGN samples should provide a better census of the underlying AGN population.
However, there has been relatively little study of the overlap and uniqueness of each AGN selection technique.  

Comparing samples selected using different techniques could provide insight into the AGNs that are missed using different selection techniques.
Similarly, it is important to understand the properties of AGNs and their host galaxies selected in each way.

Here we aim to quantify the overlap and uniqueness of various IR-AGN selection techniques compared to X-ray AGN selection.  
We use X-ray and IR data of varying depths in the Chandra Deep Field South (CDFS), COSMOS, Elias-S1 (ES1) and XMM-LSS (XMM) fields covered by the PRism MUlti-object Survey (PRIMUS) redshift survey. 
We use these datasets to investigate how the selected AGN population depends on the selection technique and the depth of the data, and we investigate the completeness, contamination, and uniqueness of X-ray 
versus IR-AGN selection.
The paper is organized as follows.
In Section~\ref{sec:data}, we present the relevant multi-wavelength datasets. 
In Section~\ref{sec:sample}, we detail the different AGN selection techniques.
In Section~\ref{sec:samplecomparison}, we compare the number densities and overlap between X-ray AGN and IR-AGN selection techniques, as a function of survey depth.
In Section~\ref{sec:contamination}, we investigate contamination of IR-AGN samples.
In Section~\ref{sec:restframe}, we compare the AGN and host galaxy rest-frame properties of X-ray and IR selection techniques. 
We discuss our results in \S\ref{sec:discussion} and conclude in \S\ref{sec:conclusions}. 
Throughout the paper we assume a standard flat $\Lambda$CDM model with $\Omega_m=0.3$, $\Omega_\Lambda=0.7$, and $H_{0}=72$~km s$^{-1}$~Mpc$^{-1}$.

\section{Data}\label{sec:data}
\begin{figure}
  \epsscale{1.1}
  \plotone{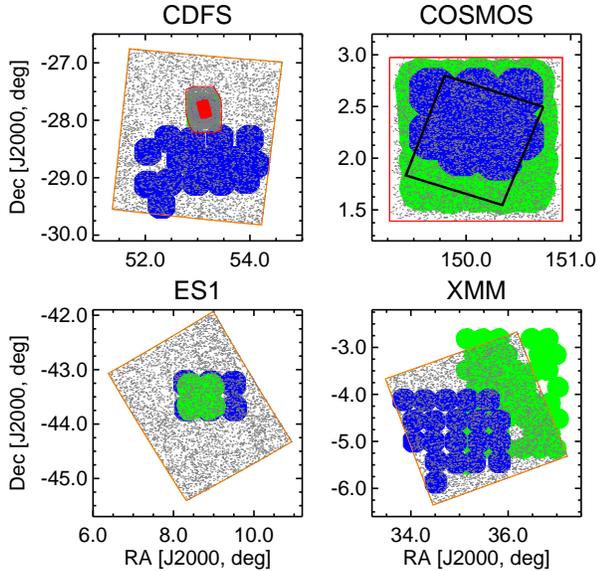}
  \caption{
Map of the IR (orange box), PRIMUS (blue), and X-ray (green) coverage in the CDFS, COSMOS, ES1, and XMM fields.
Ten thousand random objects detected in all four IRAC bands are shown with gray points.
Within CDFS the red outline shows the SIMPLE region, which has deeper IR data, and the solid red polygon shows the GOODS region, which has the deepest IR and X-ray data.
The SCOSMOS footprint is outlined in red in the COSMOS field, and the deeper X-ray footprint in the central 0.9~\degsq\ of the COSMOS field is outlined in black.
}
  \label{fig:area}
\end{figure}

For this study, we use multi-wavelength data from the CDFS, COSMOS, ES1, and XMM fields.
All of these fields have optical photometry, \spitzer\ IR imaging, spectroscopic redshifts from PRIMUS, and X-ray data from \chandra\ and \xmm.
We describe these datasets in detail in Sections \ref{sec:irdata}--\ref{sec:xraydata} below.
Figure~\ref{fig:area} shows the IRAC (orange boxes), PRIMUS (blue), and X-ray coverage (green) for each of the fields.

\subsection{\spitzer\ IR Data}\label{sec:irdata}
\begin{figure}
  \epsscale{1.2}
  \plotone{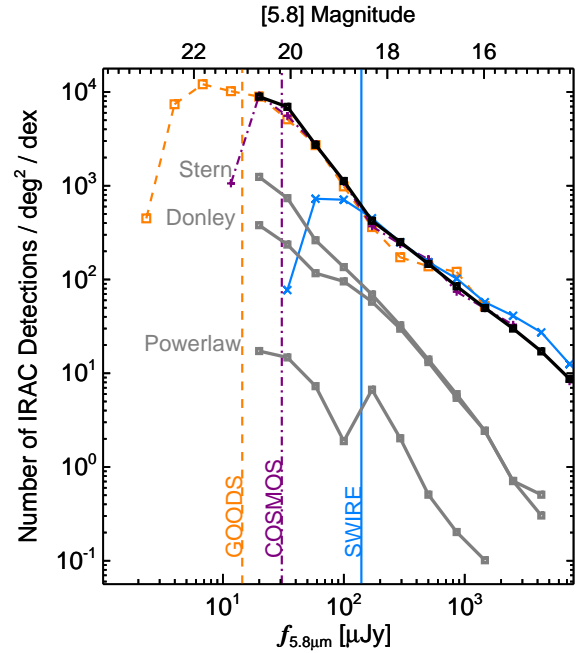}
  \caption{
Number density (number of sources per logarithmic flux bin per square degree) detected in all four IRAC channels as a function of IRAC 5.8~\um\ flux.
We show number densities for our three survey depths: \swire\ (orange squares), \cosmos\ (purple pluses), and \goods\ (light blue crosses). 
The vertical colored lines indicate our chosen 5.8~\um\ flux limit for each survey depth, which is set at a flux brighter than the observed turn over in the number counts (see Table \ref{table:fluxlimit}). 
The solid black line shows the combined number density of sources across all fields that reach our fixed 5.8~\um\ flux limits. 
We also show the \Stern, \Donley, and power-law IR-AGN total number densities from the combined sample (gray lines).
}
  \label{fig:totalnumberdensity}
\end{figure}

In the CDFS, ES1, and XMM fields, we use existing \spitzer\ IRAC imaging from Data Release 2 (DR2) of the \spitzer\ Wide-area Infrared Extragalactic Survey \citep[SWIRE; ]{Lonsdale03}.
Within the shallow CDFS-SWIRE (orange outline in Figure~\ref{fig:area}), we include deeper source catalogs within the CDFS-SIMPLE region (solid red outline) and even deeper data in CDFS-GOODS region (solid red polygon), both taken from version 3.0 of the \spitzer\ IRAC/MUSYC Public Legacy in E-CDFS (SIMPLE) survey data release \citep{Damen11}.
In the COSMOS field we use existing IRAC imaging from the S-COSMOS Survey \citep{Sanders07}.

In fields with SWIRE coverage, we adopt source catalogs provided by the SWIRE team as part of their SWIRE DR2\footnote{http://irsa.ipac.caltech.edu/data/SPITZER/SWIRE}.
These catalogs were generated by performing source detection in each of the four IRAC channels and merging the source lists (see below for more details).
For the CDFS-SIMPLE region with deeper IRAC data, we compared the measured fluxes for sources in both the SWIRE catalog and the SIMPLE catalog \citep{Damen11}. 
Based on this comparison, we find that the SIMPLE fluxes need to be scaled by \nSimpleFluxCorr, in IRAC channels 1, 2, 3, and 4, respectively to match the measured fluxes from the CDFS-SWIRE catalog. 
This multiplication factor accounts for the total flux correction applied by the SWIRE team compared to the 4\farcs0 aperture used by the SIMPLE team and a small zero point difference.
The errors are also multiplied by the above factors, and agree with the SWIRE errors after accounting for the difference in exposure times.
The SIMPLE catalog includes our deepest IRAC data in the CDFS-GOODS region.

For the S-COSMOS IRAC data, we reproduce the SWIRE source detection procedure (as outlined in the SWIRE DR2 documentation) to ensure we measure robust fluxes and errors using a consistent technique.
We downloaded the mosaic images in each IRAC channel from the NASA/IPAC Infrared Science Archive\footnote{http://irsa.ipac.caltech.edu/data/COSMOS/images/spitzer/irac/} (May 2007 release).
We use Astronomical Point source EXtractor pipeline in {\sc MOPEX} to perform source detection in the mosaic for each channel on an individual basis.
We combined the individual catalogs using the {\sc BANDMERGE} procedure in {\sc MOPEX}.
The merged sources in the final catalog are required to be detected with signal-to-noise ratio (S/N) $\ge$ 10 in channel 1 and S/N $\ge$ 5 in channel 2.
We follow the SWIRE handbook and apply a ``coverage-S/N'' flux limit, which is a factor proportional to the inverse of the square root of the coverage ($C$), the number of times that the object's location was scanned by IRAC in a given channel.
This threshold limits flux measurements to where the flux of a detected object is large compared to the statistical fluctuations in the background sky in a given channel.
Due to the deeper coverage of \spitzer\ IRAC scans in the COSMOS field, we require that the coverage threshold to be \nCosmosIracCsnr\ $\times\sqrt{4/C}$ for channels 1, 2, 3, or 4, respectively, to match the SWIRE fields.
For the majority of sources, our flux measurements are similar to those in the S-COSMOS public catalog, although the public catalog tends to have larger errors for all objects.

We combine our various fields and data sets into three ``IR surveys depths" with data of comparable depth.
\begin{enumerate}
\item \goods, the deepest area of IRAC imaging within the CDFS field.
\item \cosmos, including the S-COSMOS data \emph{and} the SIMPLE coverage in the CDFS field that reaches comparable flux limits.
\item \swire, which includes the entire area of the XMM and ES1 fields as well as the large area in the CDFS field with SWIRE coverage.
\end{enumerate}
In Figure \ref{fig:totalnumberdensity}, we show the number density of sources for each survey depth as a function of 5.8~\um\ flux.  
We require a source to be detected in all four IRAC bands, which is necessary for our analysis in this paper. 
We set a 5.8~\um\ flux limit for each IR survey depth that is just above the point where the number counts start to turn over (vertical colored lines), indicating incompleteness in the IRAC detected samples. 
Requiring a detection to the 5.8~\um\ flux limit guarantees an S/N above 2.4 and 2.1 in channels 3 and 4, respectively, with the median source having an S/N of approximately 10 in channels 3 and 4.
These flux limits are given in Table \ref{table:fluxlimit}.
Our \swire\ IR survey depth includes the entire XMM, ES1 and CDFS fields (21.55~\degsq\ in total); our \cosmos\ IR survey depth includes COSMOS and the SIMPLE region of the CDFS field (3.05~\degsq).
Our deepest IR survey depth (\goods) includes only the small GOODS region of the CDFS field (0.06~\degsq).
In the rest of this paper we adopt these limits and the corresponding area coverages when analyzing data to a given IR survey depth. 
The solid black line in Figure \ref{fig:totalnumberdensity} shows the overall number density combining all of our fields that probe above the flux limit for each IR survey depth.

\subsection{PRIMUS Data}\label{sec:primusdata}
PRIMUS \citep{Coil11} is the largest intermediate-redshift, faint-galaxy spectroscopic survey performed to date and covers $\sim$9~\degsq\ in seven different fields with existing deep multi-wavelength imaging.
We obtain low-resolution ($R \sim 40$) spectra for $\sim$300,000 objects from the IMACS instrument on the Magellan I Baade 6.5m telescope, targeting 80\% of galaxies with $i \lesssim 22$, with a statistically complete sample to a depth of $i\sim23$.
By fitting galaxy, broad-line AGNs, and stellar spectral templates to low-resolution spectra and optical ground-based photometry, we have measured $\sim$120,000 robust redshifts with a precision of $\sigma_{z}/(1+z)<0.5\%$ and a catastrophic outlier rate of $\lesssim$ 3\% ($\Delta{z}/(1+z) \ge 0.03$). 
We classify objects as galaxies, broad-line AGNs or stars based on the $\chi^2$ of the best template fits.
We derive $K$-corrections \citep{Blanton07} from the photometry.
For further details of the survey design, targeting, and data see \citet{Coil11}; for details of the data reduction, redshift confidence and precision, and completeness, see \citet{Cool13}. 

Here, we use PRIMUS redshifts between $0.2 < z < 1.2$ with high confidence quality flags ($Q \ge 3$; see \citet{Coil11}.)
In the CDFS field, the PRIMUS science observations cover part of CDFS-SWIRE but are disjoint from the CDFS-SIMPLE and CDFS-GOODS area. 
We therefore also include the PRIMUS CDFS calibration field, which overlaps with the CDFS-GOODS and CDFS-SIMPLE areas, to increase the number of deep IRAC and X-ray detected objects with PRIMUS redshifts for Section~\ref{sec:restframe}.
We thus have a total of \nPrimusRedshift\ galaxies and broad-line AGNs with high-confidence PRIMUS redshifts in the COSMOS, CDFS, ES1, and XMM fields. 
Figure~\ref{fig:area} shows the overlaps between the PRIMUS footprint and the IRAC and X-ray coverage.
Table~\ref{table:area} gives the areas of the total and overlapping regions of each data set.

\subsection{X-Ray Data}\label{sec:xraydata}
We have compiled published X-ray source catalogs based on existing \chandra\ and \xmm\ X-ray data in the COSMOS, CDFS, ES1, and XMM fields (see \citet{Aird12} for details).
In all of the fields we use the likelihood ratio matching technique \citep[e.g.,][]{Sutherland92, Ciliegi03, Brusa07, Laird09} to identify counterparts for each of the X-ray sources in the PRIMUS targeting optical photometry.
We also apply the likelihood ratio technique to assign secure IRAC counterparts to the X-ray sources in the entire area with both X-ray and \spitzer\ IRAC coverage.
Candidate counterparts are identified within 5\arcsec, although over 80\% of likelihood ratio matched IRAC counterparts are within 2\arcsec\ of their X-ray counterpart.
The likelihood ratio matching technique accounts for optical, IR, and X-ray positional uncertainties, the probability of having a counterpart with a given magnitude, and the probability of a spurious match. 
Where multiple counterparts exist, we choose the match with the highest likelihood ratio and restrict to ``secure'' counterparts with likelihood ratios $>$0.5.

Within the CDFS field, we use the \citet{Luo08} 2 Ms CDFS X-ray source catalog, one of the deepest \chandra\ survey to date, and the four overlapping 250 ks \chandra\ point source catalogs of \citet{Lehmer05}.
The 2 Ms catalog covers all of the CDFS-GOODS IRAC coverage(\nCdfsXrayCdfsGoodsArea), and part of CDFS-SIMPLE (\nCdfsXrayCdfsSimpleArea) and CDFS-SWIRE (\nCdfsXrayCdfsSwireArea) coverage, to a depth of 
\fhard{2-8}{5.5}{-17}.
The 250 ks catalogs surrounds the CDFS-GOODS field, and covers more of the CDFS-SIMPLE(\nCdfsXrayEcdfsSimpleArea) and CDFS-SWIRE(\nCdfsXrayEcdfsSwireArea) area to a depth of \fhard{2-8}{6.7}{-16}.

The COSMOS field was observed with \xmm\ \citep{Hasinger07} over the entire 2~\degsq\ to a depth of \fhard{2-10}{3}{-15}\ and with much deeper \chandra\ data reaching \fhard{2-10}{8}{-16}\ for the central $\sim$0.9~\degsq\ \citep{Elvis09}.
There are a total of \nCosmosXrayIR\ X-ray sources in the area with both IR and X-ray coverage, and \nCosmosXrayPrimus\ X-ray sources with X-ray, PRIMUS, and IR coverage; \nCosmosXrayRedshift\ of these have robust PRIMUS redshifts and classifications.

We use the \citet{Puccetti06} point source catalog in the ES1 field from a mosaic of four partially overlapping \xmm\ pointings which reached a depth of  \fhard{2-10}{2}{-15}\ covering 0.52~\degsq\ of the 0.9~\degsq\ of PRIMUS and \nElaisXrayArea\ with IR coverage.
Within the PRIMUS and IR areas (see Figure~\ref{fig:area}) there are \nElaisXrayPrimus\ and \nElaisXrayIR\ X-ray sources, respectively; there are \nElaisXrayRedshift\ X-ray sources with secure PRIMUS redshifts.

The XMM X-ray data are from the Subaru/\xmm\ Deep Survey \citep{Ueda08}, which contains 7 deep \xmm\ pointings, and the XMM-LSS X-ray survey \citep{Pierre07} which contains 45 pointings. 
The combined catalog contains X-ray sources to a depth of 
\fhard{2-10}{2}{-15}, and covers \nXmmXrayArea\ of PRIMUS and \nXmmXrayIRArea\ of the IR area.
There are \nXmmXrayPrimus\ and \nXmmXrayIR\ sources in the PRIMUS and IR overlap area, and \nXmmXrayRedshift\ objects with secure redshifts.

\section{AGN Sample Selection}\label{sec:sample}
In this section we define our samples of AGNs based on two IRAC color selection techniques and X-ray detections. 
No single selection technique can identify a complete parent sample of all AGNs; instead, we must study how the overlap between our different samples vary with the depths of the observations to determine selection biases and incompleteness effects. 
In Section~\ref{sec:sternselection}, we define our \Stern IR-AGN sample using the MIR color selection criteria from \citet{Stern05}.
We define our \Donley IR-AGN sample using the MIR color criteria from \citet{Donley12} in Section~\ref{sec:donleyselection}.
In Section~\ref{sec:powerlaw}, we discuss identification of IR-AGNs by fitting a power law to the MIR photometry \citep[e.g.,][]{Alonso-Herrero06, Polletta06, Donley07} and the reasons we choose not to use this technique in the remainder of this paper. 
We describe our X-ray-selected AGN sample in Section~\ref{sec:samplexray} and explain how we calculate completeness weights to account for the varying sensitivity of the X-ray observations.

\subsection{Stern Color Selection}\label{sec:sternselection}
Objects that are detected in all four IRAC bands are defined to be \Stern selected IR-AGNs if they have IRAC colors such that they lie within the following region in color--color space:
\begin{eqnarray}
  ({\rm [5.8]}-{\rm [8.0]}) &>& 0.6, \\
  ({\rm [3.6]}-{\rm [4.5]}) &>& 0.2\cdot({\rm [5.8]}-{\rm [8.0]}) + 0.18, \textrm{~ and}\\
  ({\rm [3.6]}-{\rm [4.5]}) &>& 2.5\cdot({\rm [5.8]}-{\rm [8.0]}) - 3.5.
\end{eqnarray}
\noindent We apply this selection technique for samples reaching our three different ``IR survey depths" defined in Section \ref{sec:irdata} above.
We identify a total of \nGoodsSternTotal, \nCosmosSternTotal, and \nSwireSternTotal\ \Stern IR-AGNs in our \goods, \cosmos, and \swire\ depth surveys, respectively. 
We show the number density of \Stern selected objects with X-ray coverage in Figure~\ref{fig:totalnumberdensity}.

\subsection{Donley Color Selection}\label{sec:donleyselection}
We also select IR-AGN samples using the IRAC color criteria presented by \citet{Donley12}.
This color-selection technique was designed to limit contamination by star-forming galaxies, especially at high redshift, but still be both complete and reliable for the identification of luminous AGNs.
We require that objects are detected in all four bands, and have IRAC colors such that they lie within the following region in IRAC color--color space:
\begin{eqnarray}
  x={\rm log_{10}}\left( \frac{f_{\rm 5.8 \um}}{f_{\rm 3.6 \um}} \right), \quad
  y={\rm log_{10}}\left( \frac{f_{\rm 8.0 \um}}{f_{\rm 4.5 \um}} \right)
\end{eqnarray}
\begin{eqnarray}
  x &\ge& 0.08 \textrm{~ and ~} y \ge 0.15\\
  y &\ge& (1.21\times{x})-0.27\\
  y &\le& (1.21\times{x})+0.27\\
  f_{\rm 4.5 \um} &>& f_{\rm 3.6 \um} \textrm{~ and ~} f_{\rm 5.8 \um} > f_{\rm 4.5 \um}, \textrm{~ and ~} \\
  f_{\rm 8.0 \um} &>& f_{\rm 5.8 \um}.
\end{eqnarray}
\noindent Like in Section~\ref{sec:sternselection}, we apply this selection method to our three different IR survey depth samples.
We identify a total of \nGoodsDonleyTotal, \nCosmosDonleyTotal, and \nSwireDonleyTotal\ \Donley IR-AGNs in our \goods, \cosmos, and \swire\ depth surveys, respectively.
The number density of \Donley selected objects with X-ray coverage is shown in Figure~\ref{fig:totalnumberdensity}.

\subsection{Power-law Selection}\label{sec:powerlaw}
We also investigated the identification of IR-AGN samples using a power-law selection technique \citet{Alonso-Herrero06, Polletta06, Donley07}.
This technique identifies IR-AGNs by fitting a power-law function to the IRAC fluxes, in contrast to the color selection techniques such as the \Stern and \Donley methods described above. 
Sources dominated by a power-law IR-AGN have a negative slope $\alpha<-0.5$ such that their $\nu F_{\nu}$ flux increases with wavelength \citep{Alonso-Herrero06}.
Following \citet{Donley07}, we identify sources with four detected IRAC flux measurements as power-law AGNs when they are well fit by a power law of slope $\alpha\le -0.5$, where $f_\nu\propto\nu^\alpha$.
We deem the power-law fit as good if $P_\chi>0.1$ (probability that the fit would have a $\chi^2$ greater than the found $\chi^2$).
We calculate the $P_\chi$ and fit parameters using the {\tt linfit} procedure in IDL.

As in Sections~\ref{sec:sternselection} and \ref{sec:donleyselection}, we apply the power-law selection technique to our \swire, \cosmos, and \goods\ IR survey depths. 
To properly simulate a shallow \swire-like survey in our \cosmos\ depth fields (COSMOS and CDFS-SIMPLE), we apply not only the 5.8~\um\ flux limits but also scale up the errors in the flux measurements to accurately represent the characteristics of the shallower data.
Specifically, we scale the errors in the fluxes by the square root of the differences in the exposure times for the \cosmos\ and \swire\ depth data. 
We verify that the median of our scaled errors matches the median error in the \swire\ depth fields.

In Figure~\ref{fig:totalnumberdensity}, we show the number density of IR-AGNs identified using the power-law selection technique as a function of 5.8~\um\ flux. 
Unlike the \Stern and \Donley samples, the number density of the power-law sample is not a smooth function of IR flux.
\citet{Donley12} show that this is mainly due to the flux errors associated with each survey. 
While we have ensured similar data reduction and consistent error estimates for the different fields and samples, the uncertainty in a flux measurement will depend on the depth of the data. 
At a given flux, deeper IR surveys find fewer power-law AGNs as the smaller flux errors can result in larger $\chi^2$ values for the power-law fit.
Thus, a source identified as a power-law AGN in a shallow survey may not be identified in a deeper survey as the more precise photometry are found to deviate from a power-law. 
Additionally the power-law sample is a very small subsample of the total X-ray population, a major problem in selecting a large statistical population of AGNs.

\citet{Donley12} suggest adopting a 10\% uncertainty floor on all flux measurements.
While this increases the number density of power-law-selected AGNs in our samples to that of \citet{Donley12}, the number density of these sources does not monotonically increase with increasing survey depth, due to the strong dependence of the power-law technique on the estimated flux uncertainty.
The power-law technique appears to be a reliable method of selecting AGNs (in that a very high fraction have X-ray counterparts), however it selects a very small subsample of the full AGN population identified either with X-ray emission or with the \Stern or \Donley AGN selection techniques using SExtractor errors.
Given the small sample size and the complex dependence on the depth of the IR data, we choose not to consider the power-law IR-AGN selection technique any further in the analysis of this paper.

\subsection{X-Ray Selection}\label{sec:samplexray}
\begin{figure}
  \epsscale{1.2}
  \plotone{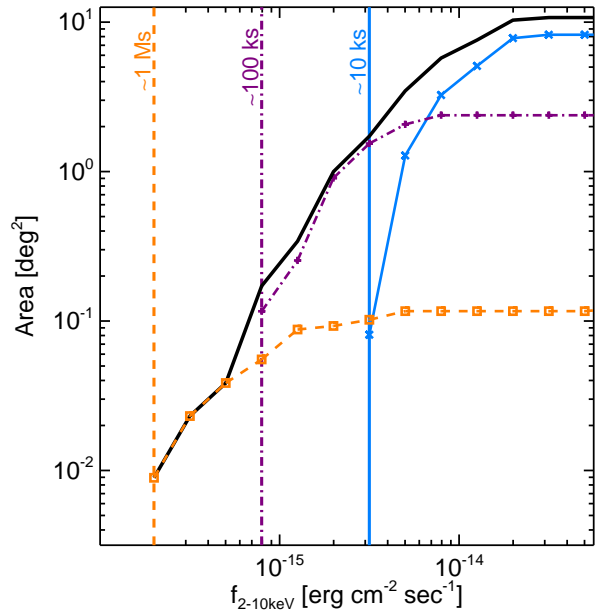}
  \caption{
X-ray area sensitivity curves that track the area which is sensitive for a given X-ray flux.
We show the individual X-ray area sensitivity curves for the $\sim$1 Ms (orange squares), $\sim$100 ks (purple pluses), and $\sim$10 ks (light blue crosses) samples.
The vertical colored lines indicate our chosen X-ray flux limit for each survey depth, which is set at the flux which we recover 10\% or more of the expected sources (see Table \ref{table:fluxlimit}). 
The solid black line shows the \All\ sample, which combines all of our X-ray fields that probe above the flux limit for each field.
}
  \label{fig:xrayarea}
\end{figure}

We define samples of X-ray-selected AGNs based on detection in the hard (2--10 keV) energy band in our compiled catalogs (see Section~\ref{sec:xraydata}).
A hard-band-selected sample includes both X-ray unabsorbed and moderately absorbed AGNs, but misses Compton-thick sources.
We do not include a soft-band-selected sample even though it probes to fainter X-ray fluxes (providing a larger sample) because it is more biased toward the detection of unabsorbed sources (with equivalent hydrogen column densities 
\Nh{<}{22}). 

Selecting in a single band also ensures we can accurately track (and correct for) the effects of variation in the X-ray sensitivity in our observations.
Within an individual \chandra\ or \xmm\ pointing, variations due to the vignetting and point-spread function of the telescope lead to variations in the X-ray flux limit that is reached across the field of view. 
Furthermore, the exposure times between different pointings can vary within an X-ray survey, leading to substantial variations in sensitivity.
The minor differences in the X-ray spectral shapes at 2$-$10 keV have a minimal effect on the sensitivity and the differences are mainly due to the X-ray telescope.
Furthermore, at the X-ray fluxes we are probing (\fhard{2-10}{\gtrsim}{-16}), the X-ray point source number counts are dominated by AGNs \citep{Georgakakis08, Lehmer12}. 
Thus, by comparing observed point source counts as a function of flux to the expected number based on the intrinsic \LogNLogS\ relation, we can estimate the fraction of our survey area that is sensitive to X-ray AGNs of a given X-ray flux. 

In Figure~\ref{fig:xrayarea}, we show X-ray area curves (the area of a survey that is sensitive to sources above a given X-ray flux) for our various fields. 
We calculate these area curves using the ratio of the number of hard X-ray-detected sources in the parent X-ray source catalog and the predicted number of sources based on the X-ray \LogNLogS\ relation of \citet{Georgakakis08}, using bins of 0.2 dex width.
We fix the area curve at the total area coverage above the lowest flux where $\ge$90\% of the predicted sources are detected to remove the effects of low source numbers at bright fluxes.
To reduce the noise due to small numbers of sources at the faintest X-ray fluxes, we set a minimum flux threshold in each X-ray survey at the point where at least 10\% of the expected number of sources are detected. 
These flux limits are shown in Figure~\ref{fig:xrayarea} and given in Table~\ref{table:fluxlimit}.

Similar to our IR observations, we group our X-ray fields into three different ``X-ray survey depths."
\begin{enumerate}
\item $\sim$1 Ms, which includes our deepest, 2 Ms X-ray data in the CDFS field.
\item $\sim$100 ks, including the E-CDFS, COSMOS, the deeper regions of the XMM field.
\item $\sim$10 ks, including ES1 field and the remainders of the COSMOS and XMM fields.
\end{enumerate}
We give nominal hard-band flux limits for each field and survey depth in Table~\ref{table:fluxlimit}.
While there are slight differences in the exact X-ray flux limit for each individual survey we show the surveys at one of the three ``X-ray survey depths'' for simplicity.
However, we note that this flux limit is only reached over a small fraction ($>10\%$) of the total area covered by our fields with a given ``X-ray survey depth''. 

Finally, we calculate an ``X-ray weight'' for each hard band X-ray detected source based on the ratio of the total area coverage to the value of the area curve at the flux of the source. 
We use these X-ray weights to correct observed number densities of X-ray sources to the intrinsic number density.
A faint X-ray source will be given a large weight to account for the small area that is sensitive to faint fluxes.
For the total X-ray sample, the X-ray weights ensure we recover the intrinsic X-ray \LogNLogS\ as measured by \citet{Georgakakis08} that was originally used to determine the area curves. 
However, the X-ray weights are also crucial when comparing between X-ray and IR-AGN samples (see Section \ref{sec:samplecomparison}).

In Section~\ref{sec:restframe}, we further limit our X-ray-detected sample to objects with an observed \Lx{>}{42}, calculated from the hard-band flux assuming a photon index of $\Gamma=1.9$.
This conservative limit ensures our sample contains objects where the X-ray flux is dominated by emission due to an AGN, rather than star formation processes in the galaxy.

When comparing the X-ray AGN sample with the IRAC AGN samples we require that both samples lie within the overlapping window functions.
Likewise for Section~\ref{sec:restframe} we require that the objects fall within the region with IRAC, X-ray, and PRIMUS coverage. 
For these studies, we recompute our X-ray weights using the overlapping areas only.


\section{Sample Comparisons} \label{sec:samplecomparison}
In this section, we examine how the total numbers of sources and the overlap between our IR-AGN and X-ray AGN samples depend on the depths of the X-ray and IRAC data. 
In Section~\ref{sec:iracxray}, we compare the IR-AGN and X-ray AGN selection techniques with respect to the bivariate IRAC and X-ray flux space.
In Section~\ref{sec:surfacedensity}, we determine the surface number densities of sources selected by the different techniques as a function of both IRAC and X-ray flux. 
In Section~\ref{sec:depth}, we investigate how the overlap between the samples depends on the IR and X-ray survey depths.

\subsection{X-Ray and IR Flux Comparison}\label{sec:iracxray}
\begin{figure}
  \epsscale{1.25}
  \plotone{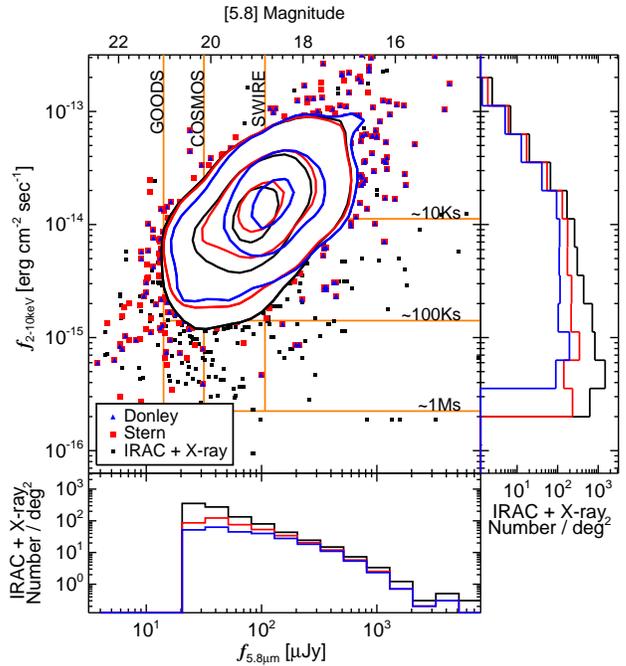}
  \caption{
Observed IRAC 5.8~\um\ and hard-band X-ray flux number-density distributions.
In the center panel, we show the sample of objects detected with four bands of IRAC and a hard-band X-ray detection.
The \Stern IR-AGN-selected sample is shown with red squares, and the \Donley IR-AGN-selected sample is shown with blue triangles.
The approximate \swire, \cosmos, and \goods\ depth IR surveys limits are shown with vertical orange lines, and approximate X-ray survey depths are shown with horizontal orange lines.
The right-side panel shows the total number of IRAC and X-ray detected of objects per logarithmic X-ray flux bin per area sensitive to that flux for the total (black), \Stern IR-AGN selected, (red) and \Donley IR-AGN selected (blue).
The lower panel shows the total number of X-ray- and IR-AGN-detected objects per logarithmic IRAC flux bin per area sensitive to that flux for the respective samples.
We include the X-ray completeness correction weights which corrects for the variation in X-ray flux limit in both side panels.
}
  \label{fig:iracxray}
\end{figure}

Our first step is to examine the distribution of sources in the bivariate space of observed IRAC 5.8~\um flux and hard-band X-ray flux.  
In the center panel of Figure~\ref{fig:iracxray} we show contours (black) that trace the distribution of all sources (across our differing survey depths) with both X-ray and IRAC detections. 
The contours contain 30\%, 50\%, and 80\% of these sources; the remaining 20\% of sources outside these contours are shown by black squares. 
We also show the distribution of sources that are also identified as IR-AGNs using the \Stern (red) and \Donley (blue) selection techniques. 
Approximate IRAC and X-ray flux limits for the different survey depths are shown by the orange vertical and horizontal lines.
We find that the majority of objects tend to be infrared bright when they are likewise X-ray bright.  
However, there is a population of objects that are infrared bright, but are X-ray faint, but very few that are X-ray bright but infrared faint.
This is expected as it suggests that it is easier to have a Compton-thick source that obscures the X-rays, but much harder to obscure the infrared photons.
Many of these infrared-bright, X-ray faint sources are not selected by either the \Stern or \Donley IR-AGN selection techniques and thus may not be AGNs or have dominate AGN components.

The right panel of Figure~\ref{fig:iracxray} shows the number density of all IRAC- and X-ray-detected objects (black), \Stern IR-AGNs (red), and \Donley IR-AGNs (blue) as a function of their observed hard-band X-ray flux.
Likewise the bottom panel shows the number density of the same samples as a function of their IRAC 5.8~\um flux.  
The number density plots in each of the subpanels are limited to include selected sources above the flux limit in each field and normalized by the area that is sensitive to that flux.
The number densities of either \Stern or \Donley IR-AGN sample do not increase toward fainter fluxes as steeply as the total number density of objects detected.

\subsection{X-Ray and IR-AGN Surface Densities} \label{sec:surfacedensity}
\begin{figure*}
  \epstrim{0in 0.9in 0.3in 0.8in}
  \epsscale{0.85}
  \plotone{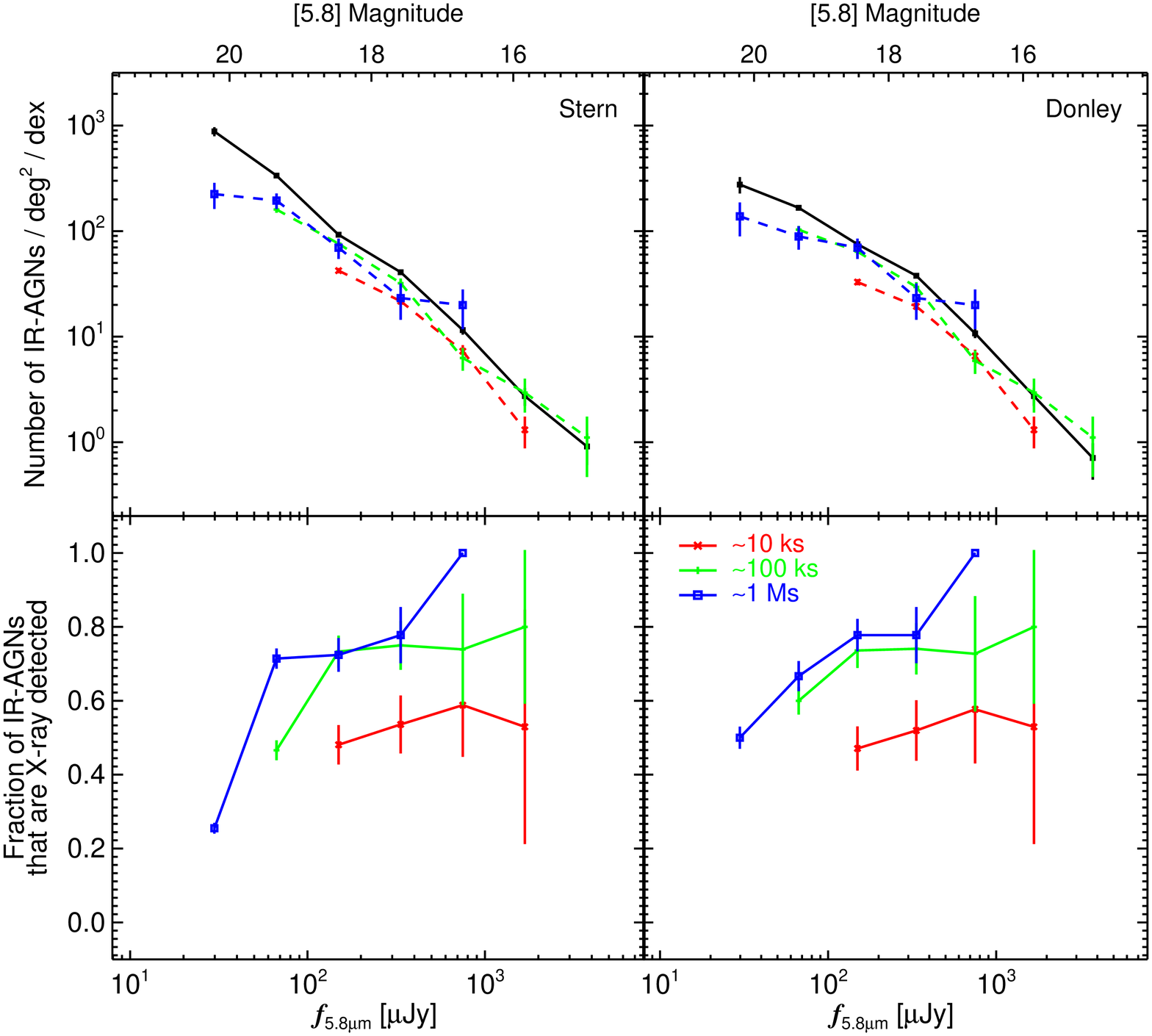}
  \caption{
IR-AGN-selected number density and fractions of combined IR-AGN samples as a function of IRAC $f_{5.8~\um}$ flux for the \Stern (left column) and \Donley (right column) IR-AGN selection techniques.
The top panels show the number density (black solid line) of the \Stern (left) and \Donley (right) IR-AGN selection method for the entire sample.
The dashed lines show the number density for individual X-ray comparison depth samples: shallow $\sim$10 ks (red), medium $\sim$100 ks (green), and deep $\sim$1Ms (blue) samples are shown with Poisson error bars.
For objects with measured X-ray fluxes we use the X-ray sensitive area derived from the X-ray completeness weights rather than the intersected IR and X-ray area.
The total sample (solid black line) may dip below an individual X-ray survey depth sample due to the X-ray completeness corrections and small number statistics, but the black line is consistent with the individual X-ray survey depth lines.
The bottom panels show the X-ray-detected fraction of IR color-selected AGNs for each of the different X-ray survey depth bins.
}
  \label{fig:squasha}
\end{figure*}

\begin{figure*}
  \epsscale{0.85}
  \epstrim{0.0in 0.8in 0.5in 1.0in}
  \plotone{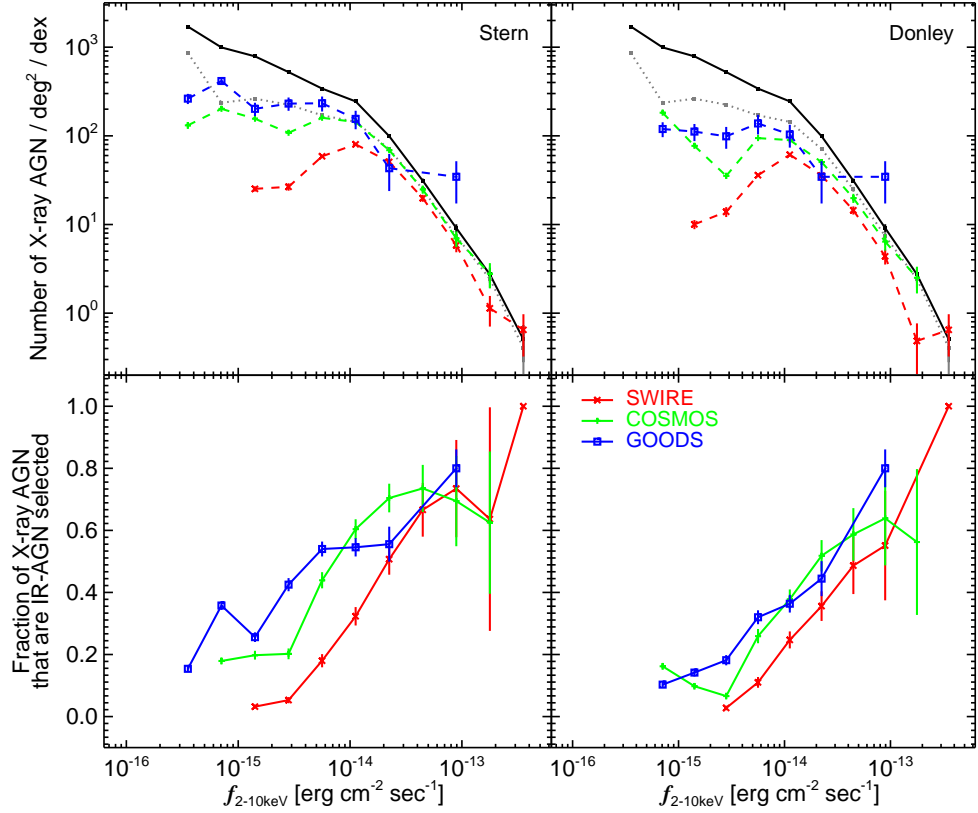}
  \caption{
Number density and fraction of the X-ray-detected sample that is selected using an IR-AGN selection technique.
The top panels show the number density of all X-ray-detected objects (black solid line) and X-ray with IRAC detected objects(gray dotted lines) as a function of hard-band X-ray. 
The dashed lines show the number density of the \Stern (left) and \Donley (right) IR-AGN samples for the shallow (\swire), medium (\cosmos), and deep (\goods) depth IR surveys.
Each of these dashed lines can be above the total number density of IR-AGNs and X-ray AGNs due to individual sample areas and small numbers at the lowest bins.
We remove bins with objects less than four objects, which happen at the largest X-ray fluxes.
The bottom panels show the IR-AGN fraction of X-ray AGN sample of the total number density for each individual IRAC depth survey.
Objects with measured X-ray fluxes have been normalized by the sensitive area for that flux measurement rather than the intersected IR and X-ray area.
}  
  \label{fig:squashb}
\end{figure*}

\begin{figure}
  \epsscale{1.2}
  \epstrim{0.2in 0.5in 0.1in 0.5in}
  \plotone{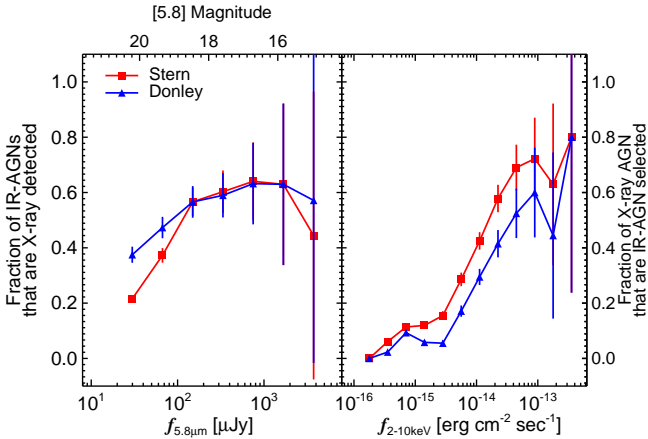}
  \caption{
X-ray-detected fraction of IR-AGNs (left panel) and the IR-AGN-selected fraction of the X-ray AGN sample (right panel) for all field depths.
The \Stern (solid red squares) and \Donley (solid blue triangles) IR-AGN-selected fraction are for all samples which contains objects above the limits for that sample.
Objects with measured X-ray fluxes have been normalized by the sensitive area for that flux measurement rather than the intersected IR and X-ray area.}  
  \label{fig:squashfraction}
\end{figure}

Our next goal is to determine how the number of sources identified as AGNs using a given selection technique varies as a function of IR flux.
In the top panels of Figure~\ref{fig:squasha}, the solid black line shows the surface number density (number per square degree per logarithmic flux interval) of sources as a function of IRAC 5.8~\um flux for the \Stern (left) and \Donley (right) IR-AGN samples. 
At the brightest fluxes (\fir{>}{100}), we combine all of the different fields to estimate the surface number density. 
At fainter fluxes we only use the fields with sufficient IR depth (see Table~\ref{table:fluxlimit} for flux limits).
For both the \Stern and \Donley IR-AGN samples the number densities increase rapidly with decreasing flux with an approximately power-law form, although the slope of the power-law may flatten below \fir{\sim}{100}\ for the \Donley sample. 
The dashed lines show the surface number densities of the \Stern or \Donley IR-AGN sample that is also detected in our X-ray data for our three different X-ray survey depths: $\sim$10 ks (red), $\sim$100 ks (green), and $\sim$1 Ms (blue).
These surface number densities are corrected for the variation in X-ray flux limit over the field (the X-ray incompleteness) by applying the weights described in Section~\ref{sec:samplexray}.

The bottom panels show the fraction of the IRAC samples that are X-ray detected for each of the X-ray survey depths. 
These fractions vary significantly as a function of both IRAC flux and the depth of the X-ray data.  
At bright IR fluxes (\fir{>}{100}) we find $\gtrsim$70\% of the IR-AGN samples are detected in X-ray data of $\sim$100 ks--1000 ks depth, whereas with the shallow X-ray data ($\sim$10 ks) this fraction is closer to 50\%. 
At fainter IR fluxes the X-ray fraction decreases, although the fall off is highly dependent on the depth of the X-ray data.
Figure~\ref{fig:squashfraction} (left) directly compares the X-ray fractions of the \Stern and \Donley IR-AGN samples as a function of IR flux combining the different X-ray survey depths. 
At bright IR fluxes $\sim$60\% of both IR-AGN samples are X-ray detected; at fainter IR fluxes a slightly higher fraction of the \Donley sources are identified in the X-ray data, although we note that the \Donley selection identifies a smaller sample of AGNs.

Our next step is to determine how the number of identified IR-AGN and X-ray AGN sources depends on the X-ray flux.
In the top panels of Figure~\ref{fig:squashb}, we show the surface number density of X-ray-selected sources as a function of the X-ray flux. 
We only include fields that probe to sufficient X-ray depth (see Table~\ref{table:fluxlimit} and Figure~\ref{fig:totalnumberdensity}) in our estimates of the surface number density at any given X-ray flux. 
We apply our X-ray weights to account for the variation in X-ray depth within a field (see Section~\ref{sec:samplexray}); only a limited fraction of the combined X-ray and IRAC area may be sensitive to a given X-ray flux, even above our nominal X-ray flux limits.
The X-ray surface number densities follow the double power-law distributions presented by \citet{Georgakakis08}.
The dashed lines show the surface number densities of X-ray sources that also satisfy the \Stern (left column) and \Donley (right column) IR-AGN selection criteria in fields reaching three different IR survey depths: \swire\ (red), \cosmos\ (green), and \goods\ (blue). 
These number densities are also corrected for variation in X-ray sensitivity within a field.
The bottom panels show the fraction of the X-ray sample that satisfy the IR selection criteria for the three IR survey depths for each of the IR-AGN samples.
At bright X-ray fluxes (\fx{\sim}{-14}), a fairly high fraction of the X-ray sources ($\sim$60\%) are identified as AGNs using either the \Stern or \Donley selection criteria, although substantially increasing the depth of the IR data does not appear to increase this fraction. 
At fainter X-ray fluxes the fraction of X-ray sources identified with the IR criteria reduces. 
Increasing the depth of the IR data from \swire\ to \cosmos\ recovers a higher fraction, but further increasing from \cosmos\ to \goods\ depths does not appear to significantly increase the IR fraction at a given X-ray flux, especially for the \Donley selection criteria.
The more restrictive \Donley selection criteria generally identifies a lower fraction of the X-ray sources as AGNs at any given X-ray flux compared to the \Stern selection (see Figure~\ref{fig:squashfraction} (right panel) for a direct comparison of the fractions with \cosmos\ X-ray and IRAC depth data.

With either of the IR-AGN selection techniques, the total number densities and the overlap between the IR-AGN and X-ray AGN samples depends on the survey depth.
In the left panel of Figure~\ref{fig:squashfraction}, we show the X-ray-detected fraction of the \Stern(red) and \Donley(blue) IR-AGN samples as a function of the IRAC flux.
For sources fainter than \fir{\sim}{100} IRAC flux, the \Stern fraction is smaller than the \Donley fraction, and very similar for sources above \fir{\sim}{100}.
Below \fir{\sim}{100}, the \Stern selection identifies a larger number of objects as AGNs than the \Donley selection (see Figure~\ref{fig:depth}), which drives this fraction down.
In the right panel of Figure~\ref{fig:squashfraction}, we show the \Stern(red) and \Donley(blue) IR-AGN fraction of the X-ray-detected sample.
The \Donley IR-AGN selection identifies a lower fraction of X-ray-detected objects for most of X-ray fluxes compared to the \Stern IR-AGN selection.

\subsection{Overlap of X-ray and IR-AGN samples} \label{sec:depth}
\begin{figure*}
  \epsscale{1.0}
  \plotone{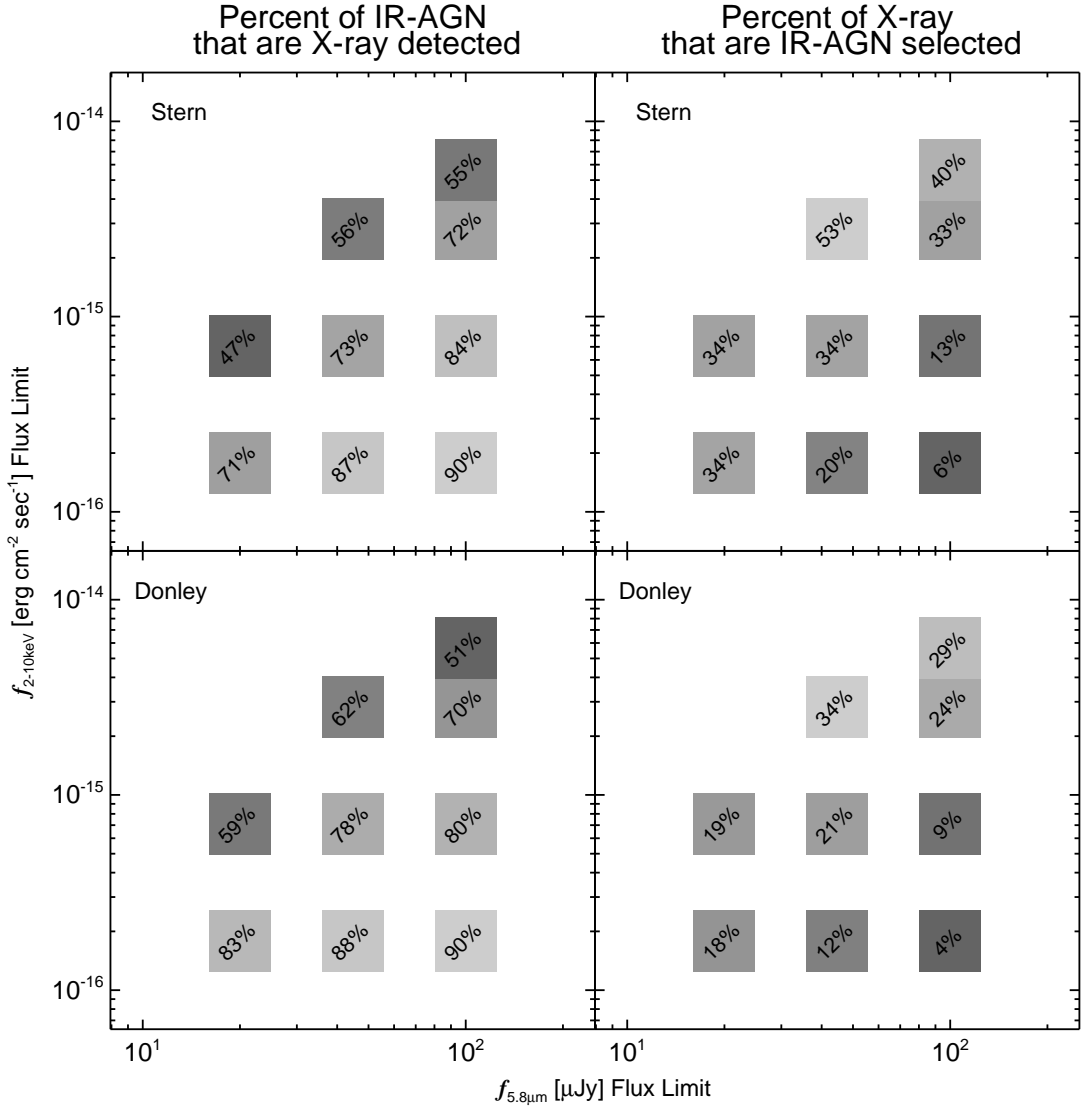}
  \caption{
X-ray-detected fraction of IR-AGN selection techniques (left column) and the color-selected fraction of X-ray AGN detections (right column) for each IR-AGN sample.
The fractions are shown at the IRAC 5.8~\um\ and X-ray flux limits with background points shown with darker gray background for lower fractions, and lighter squares for higher fractions.
The upper panels show the \Stern IR-AGN selected fractions, whereas the lower panels show the \Donley IR-AGN selected fractions. 
We include in Tables \ref{table:sternoverlap} and \ref{table:donleyoverlap} the fractions for samples with X-ray weights applied, while Tables \ref{table:sternoverlapnoweight} and \ref{table:donleyoverlapnoweight} list the raw fractions, where X-ray weights have not been applied.
}
  \label{fig:depth}
\end{figure*}

\begin{figure*}
  \epsscale{1.0}
  \epstrim{0.5in 0.8in 0.5in 0.5in}
  \plotone{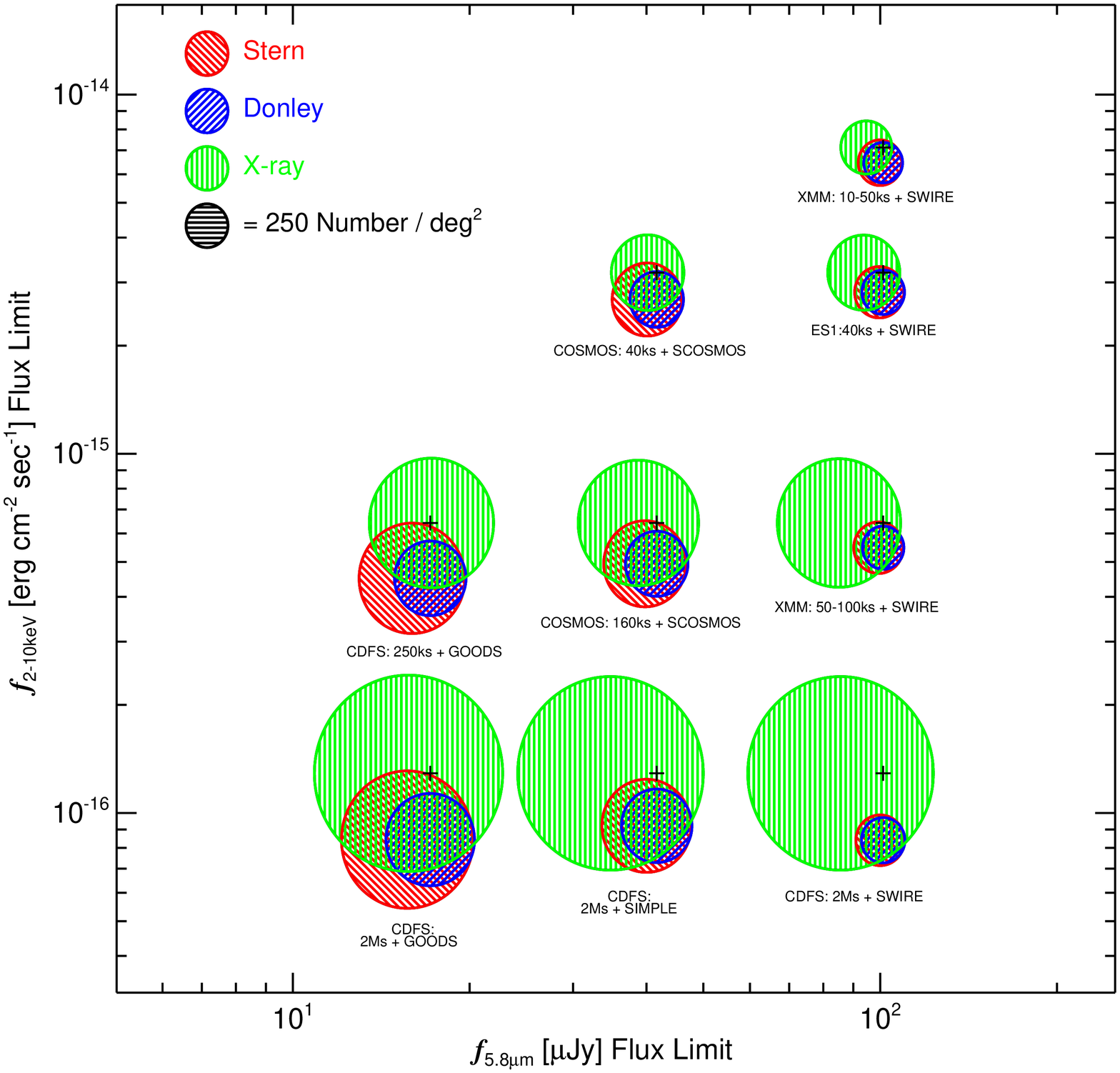}
  \caption{
Venn diagrams of the number density and overlap between the \Stern (red), \Donley (blue), and hard-band X-ray (green) selected AGN samples.
The area and overlap of each circle is proportional to the surface number density of the sample and respective sample overlap.
The different samples are located at the X-ray and IRAC flux limits shown as a black cross.
Objects with X-ray detections have been normalized by the sensitive area for that flux, and included if they fall within the X-ray and IRAC limits for that sample.
The X-ray and IRAC limit locations are plotted as a black plus at the center of the X-ray circle and \Donley selection circles, respectively. 
See Table~\ref{table:numberweight} for individual fraction values.
}
  \label{fig:venndiagram}
\end{figure*}

In this section, we examine in more detail the extent of the overlap between the IR-AGN and X-ray AGN samples and how this depends on both the IR and X-ray survey depth.
In Figure~\ref{fig:depth} (left panels), we show the fraction of the IR-AGN samples that are X-ray detected as a function of both the X-ray and IRAC flux limits for the different survey depths. 
The top-left panel shows the fraction of \Stern IR-AGNs that are X-ray detected; the bottom-left panel shows the fraction of \Donley IR-AGNs that are X-ray detected.
These fractions vary smoothly over the range of IR and X-ray flux limits between 47\% and 90\% for both \Stern and \Donley IR-AGN samples.
For comparable depth surveys in IR and X-ray (roughly the diagonal through these panels) $\sim$70\%-- 80\% of the IR-AGNs are detected at X-ray wavelengths. 
Increasing the depth of the IR data (moving to the left) reduces the fraction with X-ray detections; increasing the depth of the X-ray data (moving down) increases the fraction with X-ray detections. 
For our shallowest IR data, we find a very high fraction (90\%) of the IR-AGNs are detected using our deepest X-ray data. 
Generally, the X-ray-detected fraction of the \Donley IR-AGN samples is higher than for the \Stern IR-AGN samples, particularly for deeper surveys.

In the two right-hand panels of Figure~\ref{fig:depth}, we show the fraction of the X-ray AGN samples that are also identified as \Stern IR-AGNs (top-right panel) or \Donley IR-AGNs (bottom-right panel). 
For both selection techniques, the fraction varies smoothly between $\sim$4\% - 53\%, being smallest for shallow IR and deep X-ray samples (bottom right) and largest for deep IR and shallow X-ray samples (top left).
The \Donley IR-AGN selection technique identifies a lower fraction of the X-ray sample than the \Stern IR-AGN selection technique.
Comparing with Figure~\ref{fig:squashb}, we find that these low fractions are due to the large numbers of faint X-ray AGNs that are not selected by either IR-AGN technique, even with very deep IR data. 

In Figure~\ref{fig:venndiagram}, we use Venn diagrams to examine the surface densities and overlap of the samples selected using our three AGN selection techniques as a function of both X-ray and IR survey depth. 
The number density of the different subsamples are represented by circles located at the IR and X-ray flux limits for the different surveys.
The size of the circles are proportional to the surface number densities; the overlap between the samples are shown in Venn diagram form. 
The black cross indicates the X-ray and IR flux limits corresponding to the surveys used for each individual Venn diagram.
The number density of X-ray AGNs (green circles) increases rapidly as the X-ray depth increases (moving down in Figure~\ref{fig:venndiagram}). 
Likewise, for deeper IRAC surveys we find a larger number density of both \Stern and \Donley IR-AGN samples.
The overlap between the AGN samples varies substantially depending on the IR and X-ray survey depth.
X-ray selection identifies a large population of sources at all survey depths that are not identified with the IR-AGN selection techniques, even with extremely deep IR data. 
The \Donley IR-AGN selection technique generally identifies a subset of the \Stern IR-AGN sample for all depths. 
At our shallowest IR survey depth (\swire: right-hand side of Figure~\ref{fig:venndiagram}), the IR-AGN samples overlap almost completely, while in deeper IR surveys an additional population is identified using the \Stern selection criteria.
At the shallowest IR survey depth, a high fraction of the IR-AGNs are X-ray detected: up to 90\% with our deepest X-ray data.
Deeper IR data identifies additional IR-AGNs, including populations that are not identified at X-ray wavelengths.


\section{Contamination of IR-AGN selection}\label{sec:contamination}
In this section, we investigate whether either of the IR-AGN samples are subject to contamination from non-AGN galaxies.
In Section~\ref{sec:redshift}, we examine the redshift distributions of the IR-AGN and X-ray AGN samples (using spectroscopic redshifts from PRIMUS at $z < 1.2$) and find clear evidence for contamination of the \Stern IR-AGN sample at specific redshifts.
In Section~\ref{sec:templates}, we use MIR templates to characterize the contamination from non-AGN host galaxies that fall into the \Stern IR-AGN selection criteria compared to the \Donley IR-AGN selection criteria and X-ray detected AGNs at $z < 1.2$.
Finally in Section~\ref{sec:nonredshifts} we extend our template-based analysis to higher redshifts (z $\sim$ 2--3), outside the range covered by the PRIMUS data, to investigate the contamination of the IR-AGN samples by star-forming galaxies at these redshifts.
In this section, we focus on the templates and possible galaxy contamination in the \Stern and \Donley IR-AGN samples and leave discussion of contamination as a function of survey depth to Section~\ref{sec:discussion}.

\subsection{Investigating Contamination with PRIMUS Redshifts} \label{sec:redshift}
\begin{figure}
  \epsscale{0.85}
  \epstrim{0.5in 0.3in 0.5in 0.0in}
  \plotone{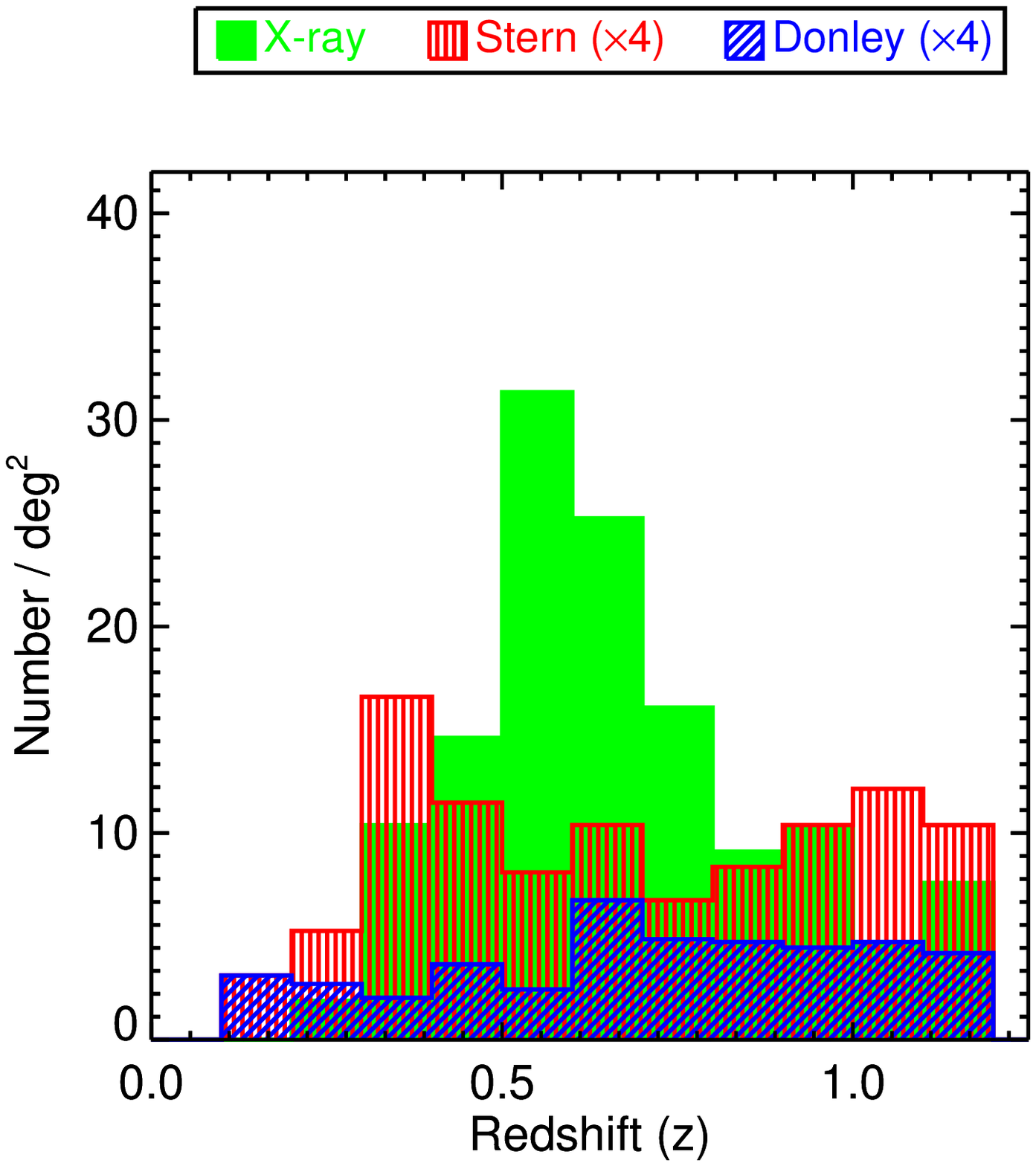}
  \caption{
Redshift number-density distributions for the X-ray (green solid), \Stern (red hatched) and \Donley (blue hatched) selected samples for all objects with secure PRIMUS redshifts in all fields.
The \Stern and \Donley IR-AGN distributions have been scaled up by a factor of four to easily compare their distributions to the X-ray sample.
}
\label{fig:redshifthist}
\end{figure}

\begin{figure*}
  \epsscale{1.2}
  \epstrim{0.4in 0.0in 0.5in 0.1in}
  \plotone{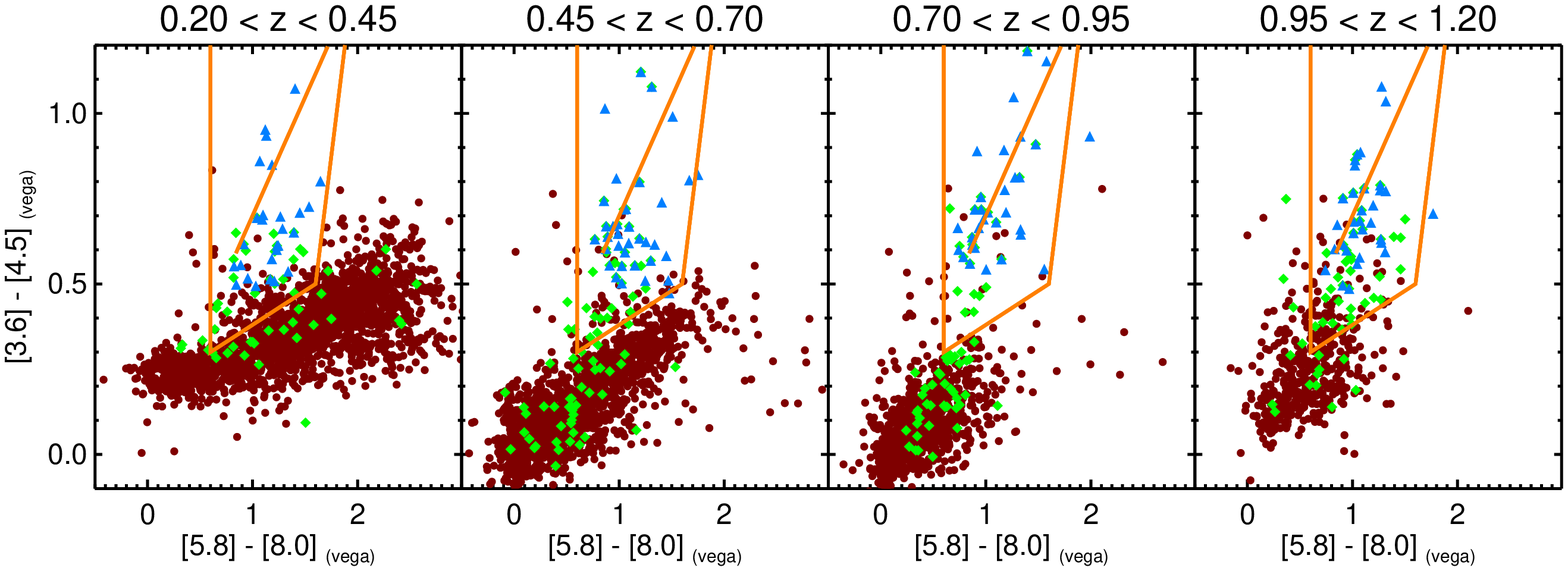}
  \caption{
IRAC color--color space for sources with PRIMUS redshifts in four redshift bins between $0.2 < z < 1.2$.
In each panel, the \Stern wedge is outlined in orange with the power-law $-0.5 < \alpha < -3$ line also shown in orange within the wedge.
The \Donley IR-AGN sample (blue triangles) and X-ray AGN sample (green diamonds) are shown above all other objects (dark red circles) in each redshift bin.
We find that the bimodal redshift peaks of the \Stern IR-AGN sample are due to a large number of objects entering the bottom of the wedge at both $z \sim 0.3$ and $z \sim 1.1$.
}
\label{fig:redshift}
\end{figure*}

In this section we compare the redshift distributions of our three AGN samples to investigate whether the selection techniques identifying the same underlying AGN population.
In Figure~\ref{fig:redshifthist}, we show the number per square degree of \Stern IR-AGN (red hatched), \Donley IR-AGN (blue hatched) and X-ray AGN (green solid) selected objects with PRIMUS redshifts.
We restrict our samples to objects with secure PRIMUS redshifts between $0.2 < z < 1.2$, where PRIMUS provides reliable redshifts for sources where the optical light is dominated by either the AGN or galaxy light and there is enough volume to construct a statistical sample.
For the X-ray detected sample we ensure that the sample is not contaminated by X-ray detected star-forming galaxies by restricting to sources with \Lx{>}{42}.
We apply the weights calculated in Section~\ref{sec:samplexray} to correct for the varying depth of the X-ray data but we do not correct for the fraction without PRIMUS redshifts.
Requiring a redshift reduces our AGN sample sizes and imposes an apparent limiting optical magnitude brighter than $i \sim 23$.
The fraction of sources in our AGN samples that are brighter than this optical limit varies from $\sim$30\% to $\sim$70\%, depending on the depth of the corresponding IR or X-ray data.
Of the sources targeted by PRIMUS, the fraction that have a high-quality redshift varies from $\sim$75\% to $\sim$95\%, depending on the depth of the IR or X-ray data, where the fraction is higher for shallower fields.
This is comparable to the overall PRIMUS redshift success rate.
We find comparably low catastrophic redshift outlier rates for the AGN samples compared to the full PRIMUS galaxy sample.
The \Stern and \Donley distributions have been scaled up by a factor of four to show their relative redshift distributions compared to the completeness-corrected X-ray sample.

The \Donley IR-AGN sample and X-ray AGN sample number-density distributions peak around $z \sim 0.6$, which matches the peak of the entire PRIMUS galaxy sample, although the \Donley distribution does not falloff as quickly for higher redshifts compared to the X-ray AGN sample.
We find that the redshift distribution of the \Stern IR-AGN sample has an unexpected shape with two peaks, at $z \sim 0.3$ and $z \sim 1.1$, which differs significantly from that of the \Donley IR-AGN and X-ray AGN samples or the overall PRIMUS galaxy sample.
To investigate this unusual distribution, in Figure~\ref{fig:redshift} we present IRAC color--color plots for the entire sample with IRAC detections in all four bands separated into four distinct redshift ranges. 
We overplot the \Donley IR-AGN (blue triangle) and X-ray AGN (green diamond) selected samples.
Generally, the \Donley IR-AGN sample is a subset of the \Stern IR-AGN sample, whereas the X-ray AGN sample includes objects both inside and outside of the \Stern color wedge.
The \Stern IR-AGN sample includes many other objects that are neither X-ray detected nor \Donley selected.
We see that the majority of these objects enter in at low redshifts ($z \sim 0.3$, left-most panel) and high redshifts ($z \sim 1.0$, right-most panel).
These objects mostly enter into the wedge at low ([3.6]--[4.5]) color 
and appear to be consistent with scatter from the larger galaxy population (the dark red points in Figure~\ref{fig:redshift}).
We thus conclude that the \Stern IR-AGN sample suffers from significant contamination by normal, non-AGN galaxies at redshifts $z \sim 0.3$ and $z \sim 1.1$, resulting in the bimodal peaks in the redshift distribution seen in Figure \ref{fig:redshifthist}.

\subsection{Comparison with IR SED Templates} \label{sec:templates}
\begin{figure*}
  \epsscale{1.2}
  \epstrim{0.4in 0.0in 0.5in 0.1in}
  \plotone{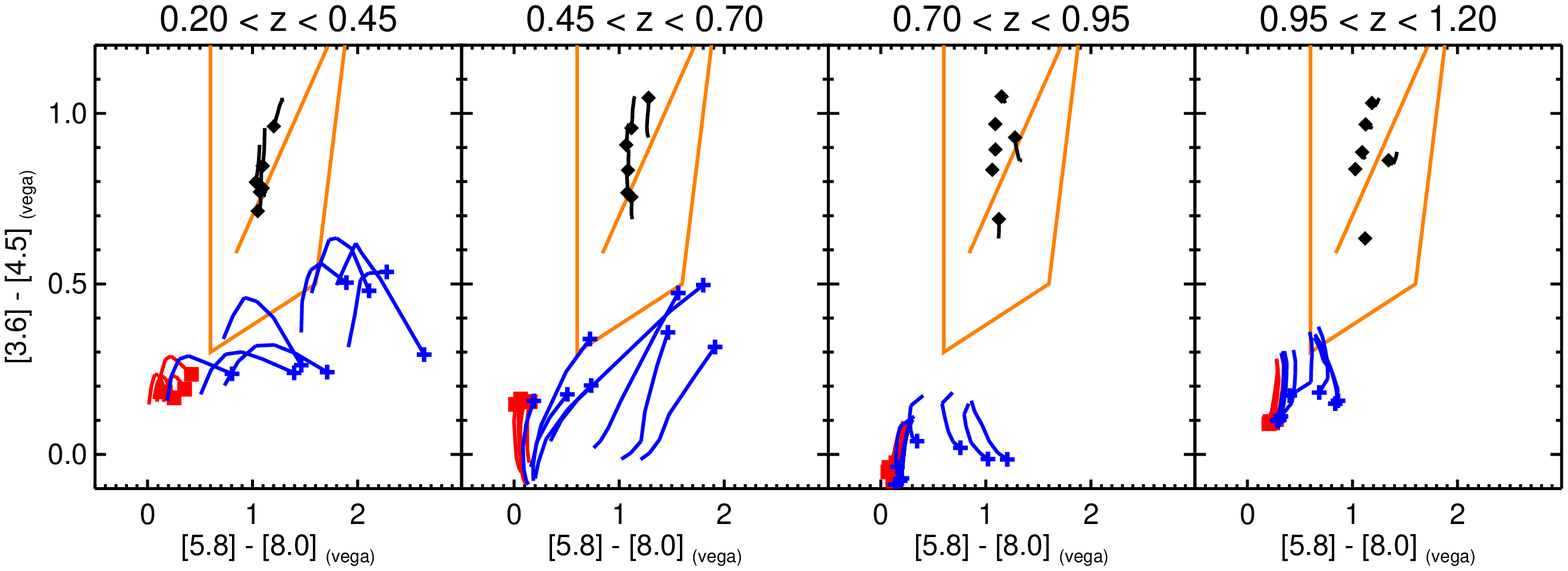}
  \caption{
IRAC color--color spaces for galaxies and AGN templates in four redshift bins between $0.2 < z < 1.20$ to match the PRIMUS redshift sample.
In each panel we show the redshifted colors of templates in that redshift bin (lines).
The \Stern wedge is outlined in orange with the power-law $-0.5 < \alpha < -3.0$ line also shown in orange within the wedge.
Quiescent galaxies (red square), star-forming galaxies (blue cross) and QSO/AGN (black diamond) redshift template tracks begin in each redshift bin at the filled symbol, and continue for the extent of that redshift bin. 
Generally, we find the \Donley and \Stern IR-AGN techniques select QSO/AGN templates.
The \Stern IR-AGN technique additionally selects star-forming templates at both $z \sim 0.3$ and $z \sim 1.1$.
Quiescent galaxy templates are not selected by either at these redshifts.}
  \label{fig:template}
\end{figure*}

In this section, we investigate the trends in the previous section using empirical galaxy and AGN SED templates.
Following \citet{Donley08, Donley12}, we redshift the \citet{Polletta08} IR templates and examine the changes to the observed IRAC colors. 
The \citet{Polletta08} IR template set consists of four quiescent galaxy templates, eight star-forming galaxy templates, and six AGN templates, based on the observed SED of a representative sample of local galaxies and AGNs.
We convolve the redshifted SED with each IRAC filter to predict the observed IRAC colors.
In Figure~\ref{fig:template}, we show the predicted colors in four redshift bins for all the star-forming galaxy (blue dots), quiescent galaxy (red squares) and AGN (black crosses) templates.
The lines track the change in color within the redshift bin for a given template.

Generally, star-forming templates change the most in this color--color space, due to the polycyclic aromatic hydrocarbon (PAH) features in the SED.
They start off with a very red ([5.8]--[8.0]) color that moves blue-ward due to the 8.0~\um\ PAH feature redshifting out of the 8.0~\um\ IRAC filter.
The vertical ([3.6]--[4.5]) motion of star-forming templates is dominated by the 3.3~\um\ PAH feature moving from the 3.6~\um\ to the 4.5~\um\ IRAC filter\footnote{The normalization of the 3.3~\um\ feature in the templates may be too high relative to the continuum, which results in redder colors than are seen in the PRIMUS sample.
However, analyzing the detailed correspondence between star-forming galaxy templates and the IRAC colors of real galaxies is beyond the scope of this paper.}.
Together, both of these features can move the star-forming templates into the \Stern wedge at both low redshift ($z \sim 0.3$) and high redshift ($z \sim 1.1$). 
The star-forming templates that come into the \Stern wedge are not selected by the \Donley IR-AGN criteria.

For quiescent templates (red squares), we find a much simpler motion through the IRAC color--color space.
Comprised of older stars, these templates are generally dominated by the Rayleigh--Jeans tail with an IRAC power-law slope of $\alpha \sim$ 2 located close to the origin.
The SEDs are generally featureless, and the motion is dominated by the 1.6~\um\ stellar bump which redshifts into the 3.6~\um\ IRAC filter.
These templates are generally excluded from either the \Donley or \Stern IR-AGN samples at these redshifts.

The predicted AGN template colors also show very little motion in this space.
Their featureless, red power-law ($\alpha\sim\text{ --}0.5 - \text{ --}3$) SEDs tend to lie close to the power-law line. 
Both \Donley and \Stern IR-AGN selection techniques pick up these AGN/quasar (QSO) templates easily at $z < 1.2$.

The behaviors of the star-forming, quiescent, and AGN templates are consistent with the trends seen in the data in the previous section and explains the peaks in the \Stern IR-AGN redshift distribution.
The \Stern IR-AGN sample is contaminated by star-forming galaxies at both low ($z \sim 0.3$) and high ($z \sim 1.1$) redshifts.
The \Donley IR-AGN selection technique identifies AGN dominated templates at $z < 1.2$.

The addition of a small AGN contribution(5\%--10\%) may move star-forming and quiescent SEDs into the \Stern wedge, but such sources are not selected by the \Donley criteria \citep[e.g.,][]{Donley12}.
Thus the \Stern selection may identify a more complete sample of AGNs but it is hampered by a larger contamination by star-forming galaxies at particular redshifts.

\subsection{Extending Templates to Higher Redshifts}\label{sec:nonredshifts}
\begin{figure}
  \epsscale{1.2}
  \epstrim{0.0in 0.2in 0.0in 0.8in}
  \plotone{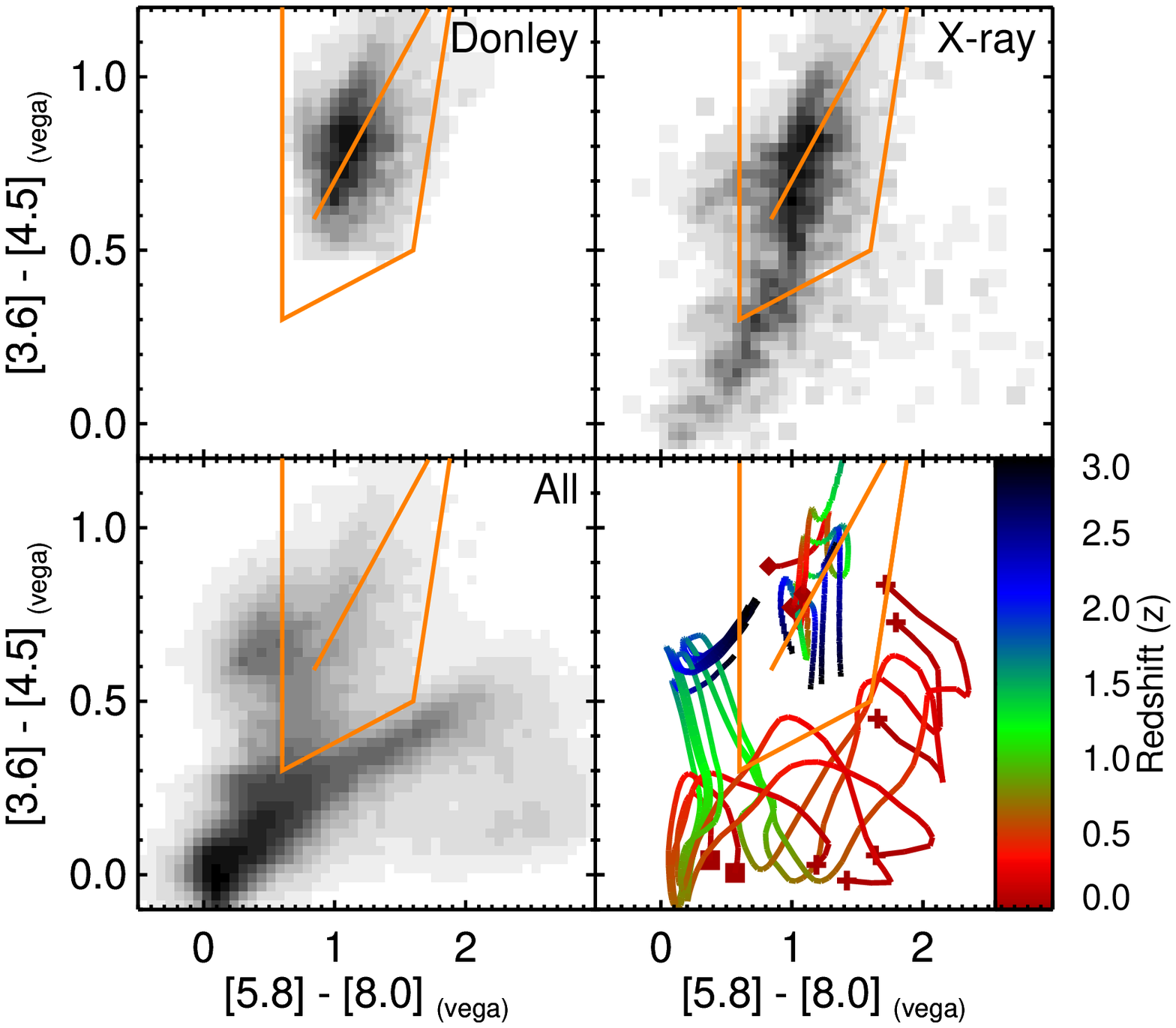}
  \caption{
Comparison of the \Donley IR-AGN and X-ray AGN samples to the entire IR-AGN sample and the IR SED templates in IRAC color--color space.
In the \Donley (top left), X-ray (top right), and \All\ (bottom left) panels, we indicate the density of IRAC-detected objects by the grayscale color.
We use a linear grayscale where black corresponds to the maximum density of objects in a given panel.
The \Stern wedge is outlined in orange with the power-law $-0.5 < \alpha < -3.0$ line also shown in orange within the wedge.
In the bottom-right panel, we show the IRAC colors of the templates from $z = 0$ to $z = 3$ (indicated by the color coding) for star-forming galaxies(cross at $z = 0$), quiescent galaxies (square at $z = 0$), and AGN (diamond at $z = 0$).  
The templates indicate that the \Stern IR-AGN sample is contaminated by galaxies at high redshifts($z\gtrsim2$), which is confirmed by our observational dataset (the over-density in the \All\ panel to the left of the \Stern wedge).
}
\label{fig:colorcolor}
\end{figure}

In this section, we extend the galaxy and AGN templates to higher redshifts ($z \ge 1.2$) beyond the range of PRIMUS to investigate the contamination of the IR-AGN samples.
We track the templates out to $z = 3$ and compare to the full sample with IRAC detections thus including objects without spectroscopic redshifts.
In Figure~\ref{fig:colorcolor}, we show the density of objects in IRAC color--color space for the \Donley IR-AGN (top-left panel), and X-ray AGN (top-right panel) samples, and for the entire sample (\All) with IRAC detections (bottom-left panel).
In the bottom-right panel we show the behavior of the templates from $z = 0$ to $z = 3$.

The \Donley IR-AGN sample (top-left panel) is centered around the power-law locus with the vast majority of the sample lying within the \Stern wedge.
It spans the range of AGN/QSO templates (seen in the lower-right panel) over the entire redshift range ($z = 0-3$).

In comparison the X-ray AGN sample (top-right panel) not only spans the range of the \Donley sample, but also extends beyond the \Donley IR-AGN sample and the \Stern wedge.
At ([3.5]--[4.5]) $\sim$ 0.75 and ([5.8]--[8.0]) $>$ 0.5 (left of the \Stern wedge) there is a spur of X-ray sources that do not have a secure PRIMUS redshift and are not selected with the \Donley IR-AGN method.
Comparing this spur with the template track panel, this region of objects appears to match both high-redshift ($z \gtrsim 2.5$) quiescent galaxy (square) and star-forming galaxies (cross).
It is unclear whether these objects contain an X-ray bright AGN and are missed by the \Donley and \Stern methods, or are merely X-ray-detected high-redshift star-forming galaxy contaminants.
If we assume that these objects have a $z = 3$ redshift, the hard band X-ray luminosity would be distributed around \Lx{>}{44.6}, which is suggestive that they are highly luminous AGN and not high-redshift star-forming galaxy contaminants.
The objects below the \Stern wedge generally match quiescent galaxy templates with $z < 1.5$ and star-forming templates with $0.7 < z < 1.5$.
This population has been shown in Section~\ref{sec:redshift} to be low redshift galaxies with PRIMUS redshifts ($z \sim 0.7$) containing genuine AGN (\Lx{>}{42}) that are not identified by either IR-AGN selection technique.

The small population of X-ray-detected sources with very red colors (([5.8]--[8.0]) $>$ 2) to the bottom right of the \Stern wedge are generally identified as low-redshift ($z \lesssim 0.4$) star forming galaxies with low X-ray luminosities (\Lx{<}{42}) and thus the X-ray emission may be attributed to star formation rather than AGN activity.
This population is easily identified and not selected by either the \Stern or \Donley IR-AGN selection techniques.

For the entire sample of IRAC-detected objects (\All, bottom left panel), the bulk of the objects lie below the \Stern wedge, concentrating at colors of ([3.6]--[4.5]) $<$ 0.0 and ([5.8]--[8.0]) $>$ 0.1.
The \All\ sample contains four-bands IRAC detected objects from all of our fields.
The colors of these objects correlate with predicted star-forming and quiescent galaxy templates colors at $0.7 < z < 1.5$
There is also a large density of sources to the left side of the \Stern wedge, like the $z > 2.5$ spur of X-ray detected objects discussed earlier.
The colors of these objects match both quiescent and star-forming galaxies SED templates at high redshift ($z \sim 2.5$).
A low fraction ($\sim$5\%) of these objects are high-redshift high-luminosity X-ray-detected AGNs, but the majority of these objects are not X-ray detected.
These objects are not identified by the \Donley selection technique, but extend into the \Stern wedge contaminating it with high-redshift galaxies.

Some of these sources in the \Stern wedge could have a low-luminosity AGN, but the AGN signature is being masked in the MIR by the dominant galaxy light.
Such AGNs could be identified using other techniques (e.g., X-ray, or via their optical spectra), but the true AGNs are probably a minority of the contaminating population.
The dominant galaxy light suggest that they are not generally broad-line QSOs, but could host narrow-line AGNs.

In summary, the \Stern IR-AGN technique selects a large population of AGNs but is contaminated by low-, intermediate-, and high-redshift galaxies ($z \sim 0.3$, $z \sim 1.1$, and $z \gtrsim 2.5$).
The \Donley IR-AGN selection technique generally is free of these galaxy contaminants, selecting just the AGN population.
The X-ray detected sample has colors that extend beyond the \Stern and \Donley selected samples, identifying a population of AGNs that are not detected by either IR selection technique and are not star-forming galaxy contaminants.


\section{AGNs and Host Galaxy Rest-frame Properties}\label{sec:restframe}
In this section, we investigate the AGN and host galaxy rest-frame properties of our IR-AGN and X-ray AGN samples.
In Section~\ref{sec:redshiftproperties}, we create subsamples with PRIMUS spectroscopic redshifts in order to study restframe properties.
In Section~\ref{sec:agnproperties}, we measure the AGN luminosity and hardness ratio (HR) distributions for the IR and X-ray AGN samples.
We then investigate how the X-ray luminosities vary across the IRAC color--color space.
In Section~\ref{sec:hostproperties}, we compare the host galaxy colors and stellar masses for the IR and X-ray AGN samples to investigate differences in the host galaxy properties.

\subsection{The PRIMUS Redshift Sample}\label{sec:redshiftproperties}
To investigate AGN and host galaxy rest-frame properties, we require a PRIMUS spectroscopic redshift.
This reduces our AGN sample sizes and imposes an optical apparent magnitude limit.
To be targeted by PRIMUS an object must have an optical magnitude brighter than $i \sim 23$.
The fraction of sources in our AGN samples that are brighter than this optical limit varies from $\sim$30\%--$\sim$70\% depending on the depth of the corresponding IR or X-ray data.
Of these sources that are brighter than the PRIMUS targeting limit, $\sim$20\% to $\sim$60\% were targeted in PRIMUS (not all objects to the apparent magnitude limit can be targeted due to slit collisions).
The fraction of targeted AGNs that have a high-quality redshift varies from  $\sim$75\% to $\sim$95\% depending on the depth of the IR or X-ray data, where the fraction is higher for shallower fields.
This is comparable to the overall PRIMUS redshift success rate.
Some targeted sources for which we fail to measure a redshift lie at z $>$ 1.2 where the Balmer break falls beyond the wavelength coverage of PRIMUS.
Our redshift success range in the $0.2 < z < 1.0$ range is likely to be higher.
The fraction of our IR- and X-ray-selected AGN with PRIMUS redshifts is $\sim$20\%--$\sim$70\% depending on depth of the data.
We find that the subsample with PRIMUS redshifts spans the full range of IR and X-ray fluxes in our IR and X-ray AGN samples.
In the rest of this section, we use these subsamples to compare the rest-frame properties of the IR- and X-ray-selected AGNs to an optical limit of $i \sim 23$.

\subsection{Rest-frame AGN Properties}\label{sec:agnproperties}
\begin{figure*}
  \epsscale{1.0}
  \epstrim{0.1in 0.3in 0.1in 0.4in}
  \plotone{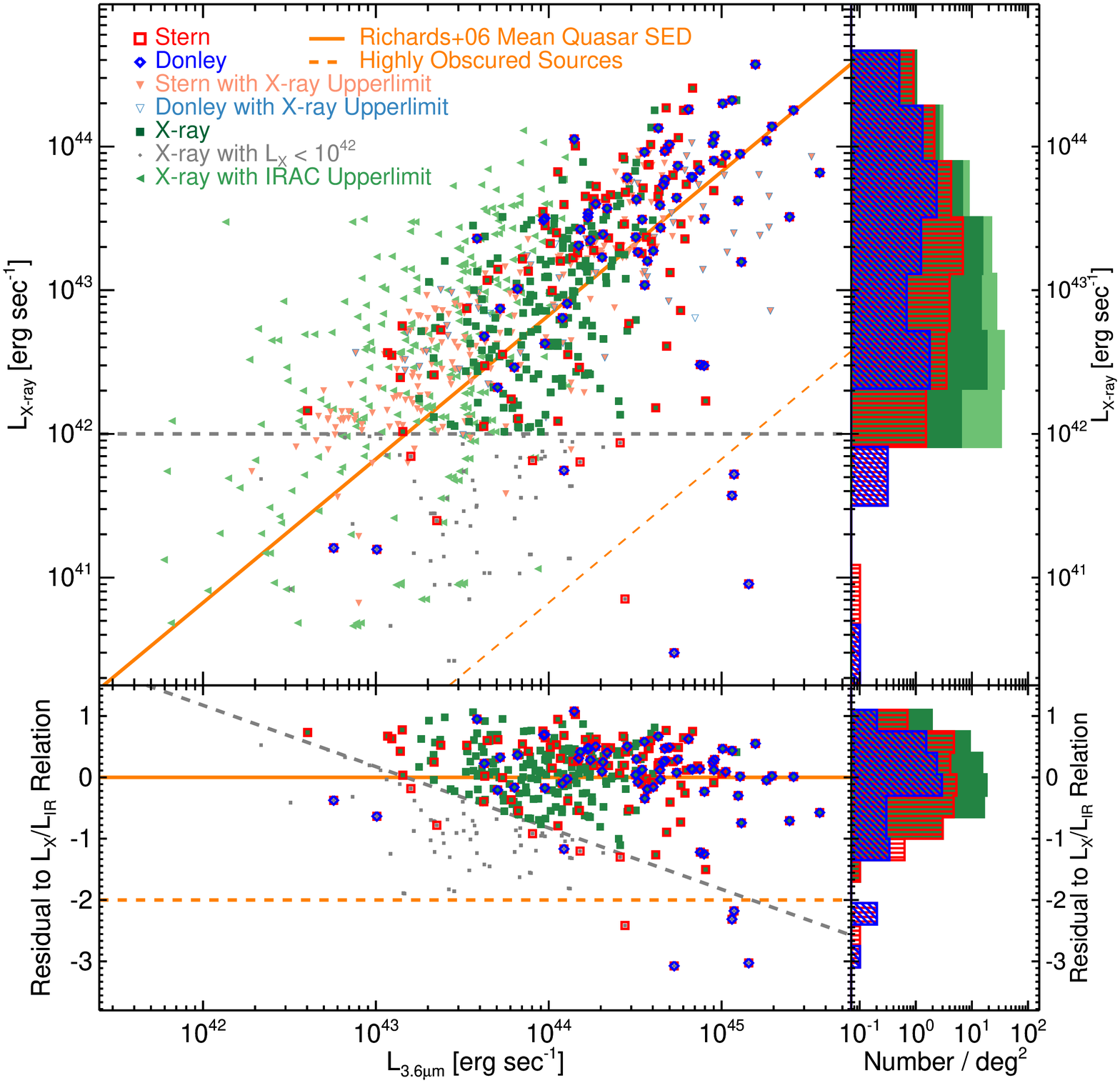}
  \caption{ 
X-ray vs. IR luminosity comparison plot (top left) for the objects that are X-ray or IRAC detected.
The \Stern selected objects are shown with a red square, the \Donley selected objects are shown with smaller blue square, and the X-ray-detected objects with \Lx{>}{42} shown in green filled squares.
We show the X-ray-detected objects with redshifts but not detected in the IRAC with light-green upper limits at the depth of the IR survey.
We show the \Stern or \Donley IR-AGNs with redshifts but not X-ray detections with light red or blue upper limits at the X-ray flux corresponding to 50\% of the expected number of X-ray sources are detected for each field.
The IR and X-ray sample luminosities are correlated; the more luminous IR sources are also the more luminous X-ray sources.
This correlation matches the \citet{Richards06} mean quasar SED, which we show as an orange solid line.
In the top-right panel, we show the X-ray luminosity number-density distribution for the sources in the previous panel.
We show the \Stern distribution with a red hatched histogram, the \Donley distribution with a blue diagonal hatched histogram, and the X-ray and (not) IR-detected sample distribution in (light) green filled (arrows) circles.
We do not show \Stern or \Donley IR-AGNs that are not X-ray detected in this histogram.
In the bottom-left panel, we show the residuals between the source luminosity and the \citet{Richards06} mean quasar SED line for the X-ray and IR detected objects.
For the distribution panels, objects with X-ray detections have been normalized by the sensitive area for that flux.
The points are shown with the same colors and styles as the top-left panel.
We find that the residuals generally scatter below the mean SED line with more sources having a higher IR luminosity than X-ray luminosity.
}
\label{fig:excess}
\end{figure*}

\begin{figure*}
  \epsscale{1.2}
  \epstrim{0.49in 0.5in 0.25in 0.25in}
  \plotone{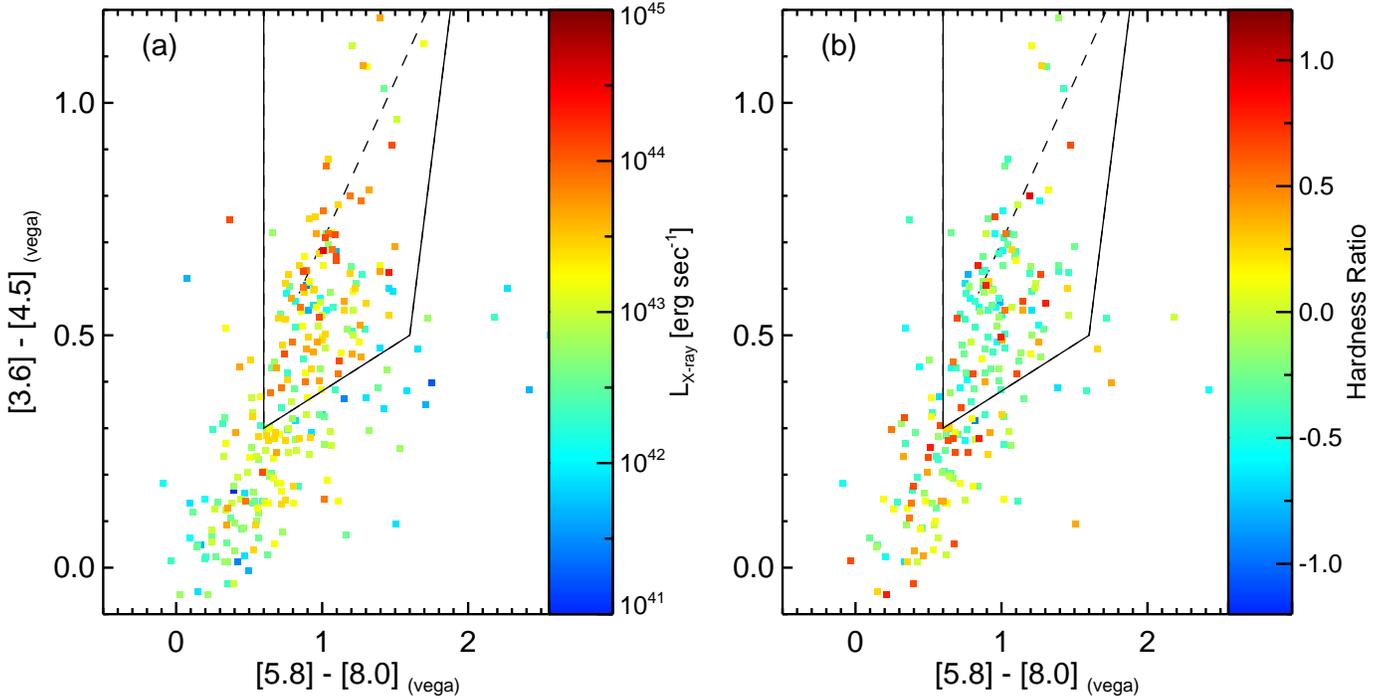}
  \caption{
(a) IRAC color--color plot of the X-ray-detected population where we scale the color of each object to the X-ray luminosity.
We show the border of the \Stern wedge as a solid black line, and the power-law locus as a dashed black line.
We find a greater than average number of high X-ray luminosity sources to be within the \Stern wedge, and more than average numbers of sources with low X-ray luminosity to be outside of the wedge.
There is a population of medium X-ray luminosity(\Lx{\sim}{43.5}) sources that are not selected by the \Stern or \Donley techniques.
(b) IRAC color--color plot of the X-ray detected population with object color scaled to the X-ray flux hardness ratio.
Generally, we find that hard sources (HR $\sim$ 1) are not preferentially found within the \Stern wedge compared to the full sample.
}
  \label{fig:luminosity}
\end{figure*}

\begin{figure}
  \epsscale{1.2}
  \epstrim{0.0in 0.3in 0.0in 0.1in}
  \plotone{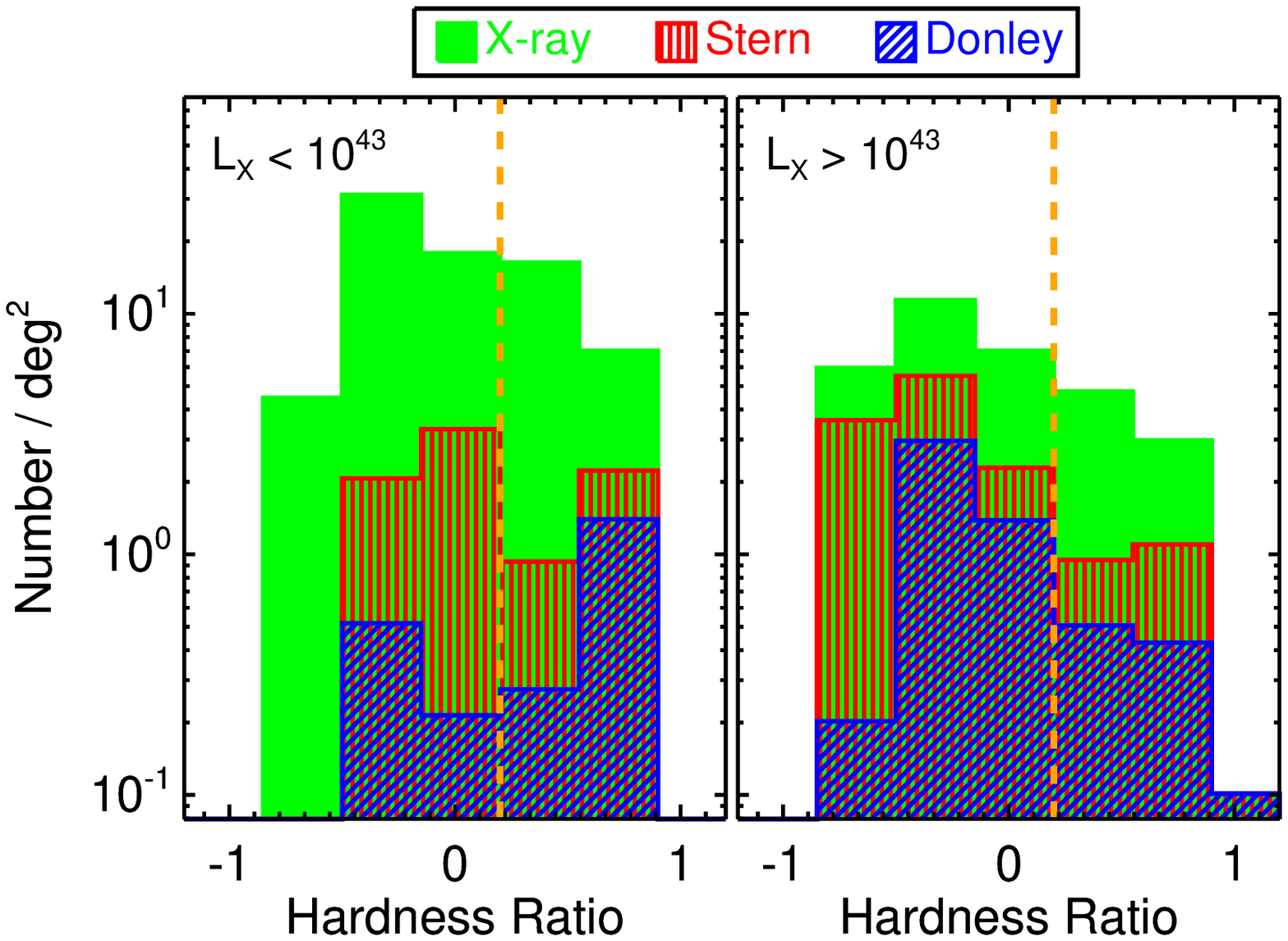}
  \caption{
Hardness ratio comparison for the \Stern (horizontal hatched red shading), \Donley (diagonal hatched blue shading), and X-ray (green filled shading) identified samples.
The two panels show the X-ray hardness ratio split up into X-ray faint sources (\Lx{<}{43}, left) and X-ray luminous sources (\Lx{>}{43}, right).
For the low X-ray luminosities  (\Lx{<}{43}), there is a hint that the IR-AGN samples identify a \textit{more} obscured population with a larger fraction of hard sources (HR $>$ 0.2); however, this is not statistically significant.
}
\label{fig:agnrestframe}
\end{figure}

\begin{figure}
  \epsscale{1.1}
  \plotone{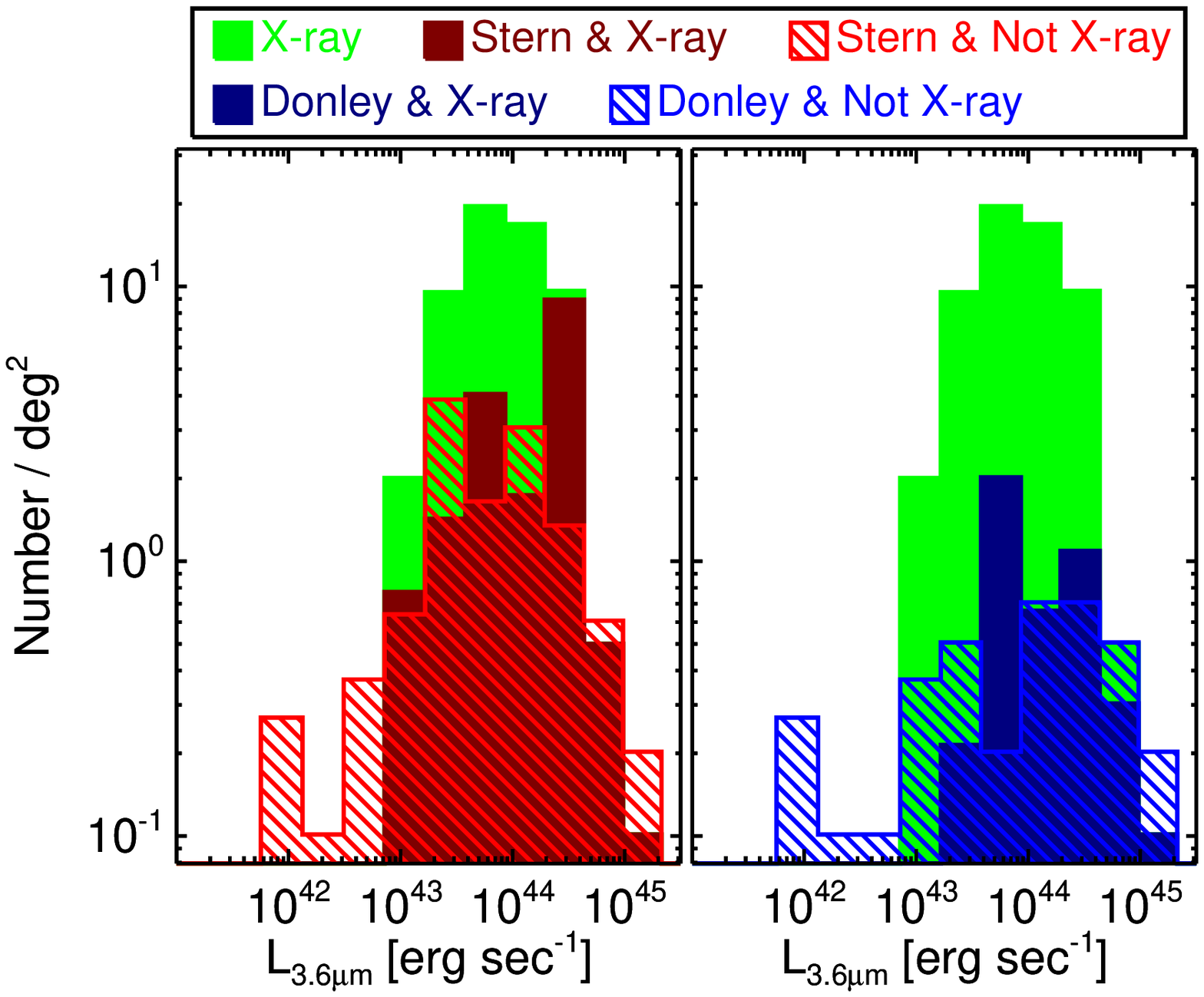}
  \caption{
3.6\um\ IR luminosity number-density distributions for sample intersections.
In both panels, we show the X-ray distribution in green.
In the left panel, we show the number density of \Stern sources that are X-ray detected (red solid) and the number density of \Stern sources that are not X-ray detected (red hatched).
In the right panel we show the number density of \Donley sources that are X-ray detected (blue solid) and the number density of \Donley sources that are not X-ray detected (blue hatched).
We do not include the X-ray completeness correction weights for either IR selection.
We find that the IR-AGN samples that are not X-ray detected have similar median luminosities compared to the X-ray detected sample.
}
  \label{fig:propcomparison}
\end{figure}

In this section, we investigate rest-frame properties related to the AGN themselves.
In Figure~\ref{fig:excess}, we compare the IR and X-ray luminosities for the AGN samples.
IR luminosities are calculated at a rest-frame wavelength of 3.6~\um\ by interpolating the log of the fluxes between the two nearest IRAC channels.
IR luminosities are calculated at a rest-frame wavelength of 3.6~\um\ in two manners: (1) interpolating the log of the fluxes using the two nearest IRAC channels, and (2) using template fit synthesizes luminosities from $K$-correct \citep{Blanton07}.
We find no significant differences in the luminosity distributions between these two methods and use the former for the remainder of the paper. 
The solid green circles indicate PRIMUS AGNs that have IR and X-ray detections and are above \Lx{>}{42}.
X-ray luminosities are calculated from the 2--10 keV fluxes, using an unobscured power-law with a photon index of $\Gamma = 1.9$.
The light green arrows indicate upper limits on the 3.6~\um\ IR luminosity for X-ray-detected AGNs that lack IR detections.
The IR luminosity upper limits are derived from the IR depth of the survey and the redshift of the object.
The X-ray luminosity upper limits for the \Stern or \Donley IR-AGNs that are not X-ray detected are estimated from the X-ray depth of each field.
We conservatively use the X-ray flux limit corresponding to the flux at which 50\% of expected X-ray sources are detected in that field.
For a more conservative 90\% detection limit, the upper limits increase by roughly a factor of 2$-$4.
Objects that are selected by either \Stern or \Donley IR-AGN techniques are indicated by red squares or blue triangles, respectively.
Generally, sources that are detected in both IR and X-ray have correlated luminosities which roughly follows the \citet{Richards06} mean QSO SED shown as a solid orange line.
The correlation does contain significant scatter with more sources scattering to much lower X-ray luminosities.
The sources that scatter to lower X-ray luminosities are found two orders of magnitude below the mean QSO SED

We find that both IR-AGN color selection techniques tend to select sources with higher IR and X-ray luminosities compared to the entire X-ray-selected AGN sample.
A weighted K-S test of the distributions of the X-ray luminosities, as shown in the top-right subpanel of Figure~\ref{fig:excess}, rejects the null hypothesis that either the \Stern or \Donley IR-AGNs are drawn from the same distribution as the X-ray selected AGN at the $> 99.9\%$ confidence level.
X-ray selection identifies a large population of AGNs with low luminosities (below \Lx{<}{43.5}) that are not identified as either \Stern or \Donley IR-AGNs.

These trends are also shown in Figure~\ref{fig:luminosity}(a), which plots the subset of X-ray AGNs detected in all four IRAC bands in IRAC color--color space, where the color of the points scale with X-ray luminosity.
In this figure we do not limit sources to be above \Lx{>}{42}.
Sources below this limit reside in the region below the \Stern or \Donley IR-AGN selection regions and are easily distinguished from IR-selected AGNs.
Low X-ray luminosity (\Lx{<}{43.5}) AGNs are scattered throughout the IRAC color--color space, whereas high X-ray luminosity (\Lx{>}{43.5}) AGNs cluster near the power-law line within the \Stern wedge.
For high X-ray luminosity (\Lx{>}{43.5}) AGNs with IRAC detections, $\sim$\nDonleyBrightLogXLum\ to $\sim$\nSternBrightLogXLum\ lie within the spaces defined by the \Donley and \Stern IR-AGN selections, respectively.

In the bottom panel of Figure~\ref{fig:excess}, we plot the residuals X-ray to IR luminosity ratio from the \citet{Richards06} mean quasar SED shown in the upper panel (solid orange line).
In this panel, we have removed the AGN not detected in the IR.
We find that there is a tail of sources with negative residuals such that their X-ray to IR luminosity ratio is below the \citet{Richards06} mean quasar SED (solid orange) line.
There are between \nSternIRObscuredFraction\ and \nDonleyIRObscuredFraction\ of \Stern and \Donley sources with residuals less than $\nExcessCut$; these sources are particularly X-ray underluminous.
These sources are likely obscured AGNs where the observed X-ray luminosity is suppressed due to absorption in the 2-10 keV X-ray band. 
We note that the IR luminosity \LIR\ may be systematically inflated by galaxy light for lower luminosity AGNs.
In the IR, the galaxy light dominates over the AGN light above an accretion rate of 1\% Eddington (see Section~\ref{sec:hostproperties}).
However, the sources that are underluminous in X-rays are very luminous in the IR (\Lir{>}{44}) such that the galaxy contribution is small relative to the AGN.

We also examine the obscuration of the X-ray AGN using HRs.
We convert the hard- and soft-band fluxes to equivalent on-axis \chandra\ count rates and calculate the HRs as $(H-S)/(H+S)$ where $H$ and $S$ are the 2-10 keV and 0.5-2 keV count rates, respectively.
Large HRs values (HR$\ge$ 0.2) indicate significant absorption in the hard X-ray band  (\Nh{\gtrsim 3 \times}{21}) at $z \sim 0.6$ \citep{Hasinger08}.
In Figure~\ref{fig:luminosity}(b), we show the HRs for the subset of X-ray AGNs detected in all four IRAC bands in IRAC color--color space and find that lower HR (unobscured) AGNs tend to fall within the \Stern wedge.
This is consistent with IR-AGN samples identifying primarily more luminous AGN, which tend to be unobscured.
Obscured AGN lie throughout the IRAC color--color space, with \nSternObscuredFraction\ in the \Stern wedge and the remainder in the region where the galaxy light dominates the IR colors.

We compare the HRs of the IR-AGN and X-ray AGN samples in a two X-ray luminosity bins to limit the large number of luminous IR-AGN sources from dominating the comparison.  
In the right panel of Figure~\ref{fig:agnrestframe}, we show the distributions of HRs for X-ray luminous sources (\Lx{>}{43}).
The \Stern IR-AGN, \Donley IR-AGN, and X-ray detected AGN samples all span a wide range of HRs.
The fraction of obscured sources ($HR > 0.2$) is comparable for all three samples.
The fraction in the X-ray AGN sample (\nXrayBrightHRObscuredFraction) is slightly larger than the fraction in the \Stern (\nSternBrightHRObscuredFraction) or \Donley (\nDonleyBrightHRObscuredFraction), although the differences are not statistically significant.

In the left panel of Figure~\ref{fig:agnrestframe}, we show the distributions of HRs for lower luminosity X-ray sources (\Lx{<}{43}).
The fraction of obscured sources (HR $> 0.2$) for the X-ray AGN sample (\nXrayFaintHRObscuredFraction) is slightly smaller than the fraction in the \Stern (\nSternFaintHRObscuredFraction) or \Donley (\nDonleyFaintHRObscuredFraction) AGN samples.
Unlike at high X-ray (\Lx{>}{43}) luminosities, there is a hint at low X-ray luminosities that the IR-AGN samples identify a \textit{more} obscured population for lower luminosity AGNs; however, there are not statistically significant differences between the HR distributions in either luminosity range.
Larger deep-IR samples are needed to confirm if this small population of obscured sources statistically differs from the X-ray sample.

Finally, in Figure~\ref{fig:propcomparison} we show the 3.6~\um\ IR luminosity number-density distribution for the IR-AGN samples that are divided into X-ray detected (solid histograms) and not X-ray detected (hatched histograms).
We also show a subset in green of the X-ray AGN sample that is detected in all IRAC four bands.
For both the \Stern and \Donley AGN samples, we do not find any significant differences in the \LIR\ distributions for the sources with X-ray detections to those without X-ray detections.
These latter sources could be missed in X-ray due to heavy obscuration, but nevertheless they have a similar distribution of IR luminosities and do not outnumber the subsample of the X-ray AGN sample detected in IRAC.
The low number density of IR-AGN sources without X-ray detections may partially be due to having relatively deeper X-ray survey data compared to the IRAC survey data.
Using the 3.6~\um\ and X-ray luminosity from the \citet{Richards06} mean quasar SED, we find that the depth of the X-ray data in our fields is typically 2--3 times deeper than the IRAC depth, for light that is dominated by luminous AGNs.
Thus, while IR-AGN samples are efficient at identifying luminous AGNs, it remains unclear if they identify a large population of obscured sources relative to X-ray AGN samples due to the differences in survey depths.

\subsection{Host Galaxy Properties}\label{sec:hostproperties}
\begin{figure*}
  \epsscale{1.15}
  \epstrim{0.45in 0.0in 0.0in 0.0in}
  \plotone{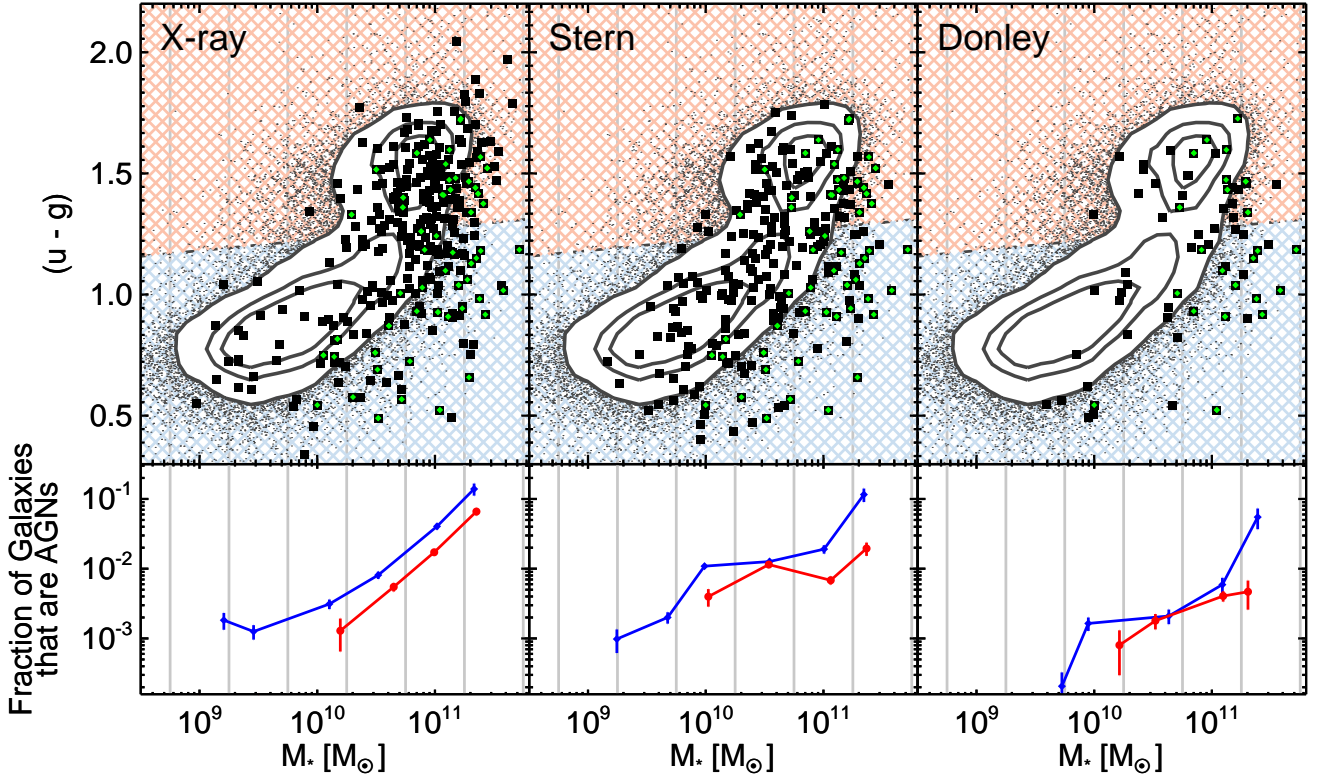}
  \caption{
Optical color vw. stellar mass diagram for the X-ray AGN (left column), \Stern IR-AGN (center column), and \Donley IR-AGN (right column) samples.
In the top panel of each column we show the rest-frame $(u-g)$ optical colors on the vertical axis and the stellar mass estimates for PRIMUS galaxies (gray points and contours) and the AGN sample (black squares).
In each panel we show the sources that are both X-ray detected and IR-AGNs selected with green diamonds.
The background colored hatching shows the approximate red and blue galaxies definition at $z = 0$ and $m_g \sim 20$.
In the bottom panel of each column, we show the fraction of detected AGNs compared to all PRIMUS galaxies at a given stellar mass for red and blue hosts.
The fraction is calculated in six stellar mass bins and is shown at the median AGN mass in that bin.
}
\label{fig:colormass}
\end{figure*}

\begin{figure*}
  \epsscale{0.99}
  \epstrim{0.25in 0.6in 0.8in 0.8in}
  \plotone{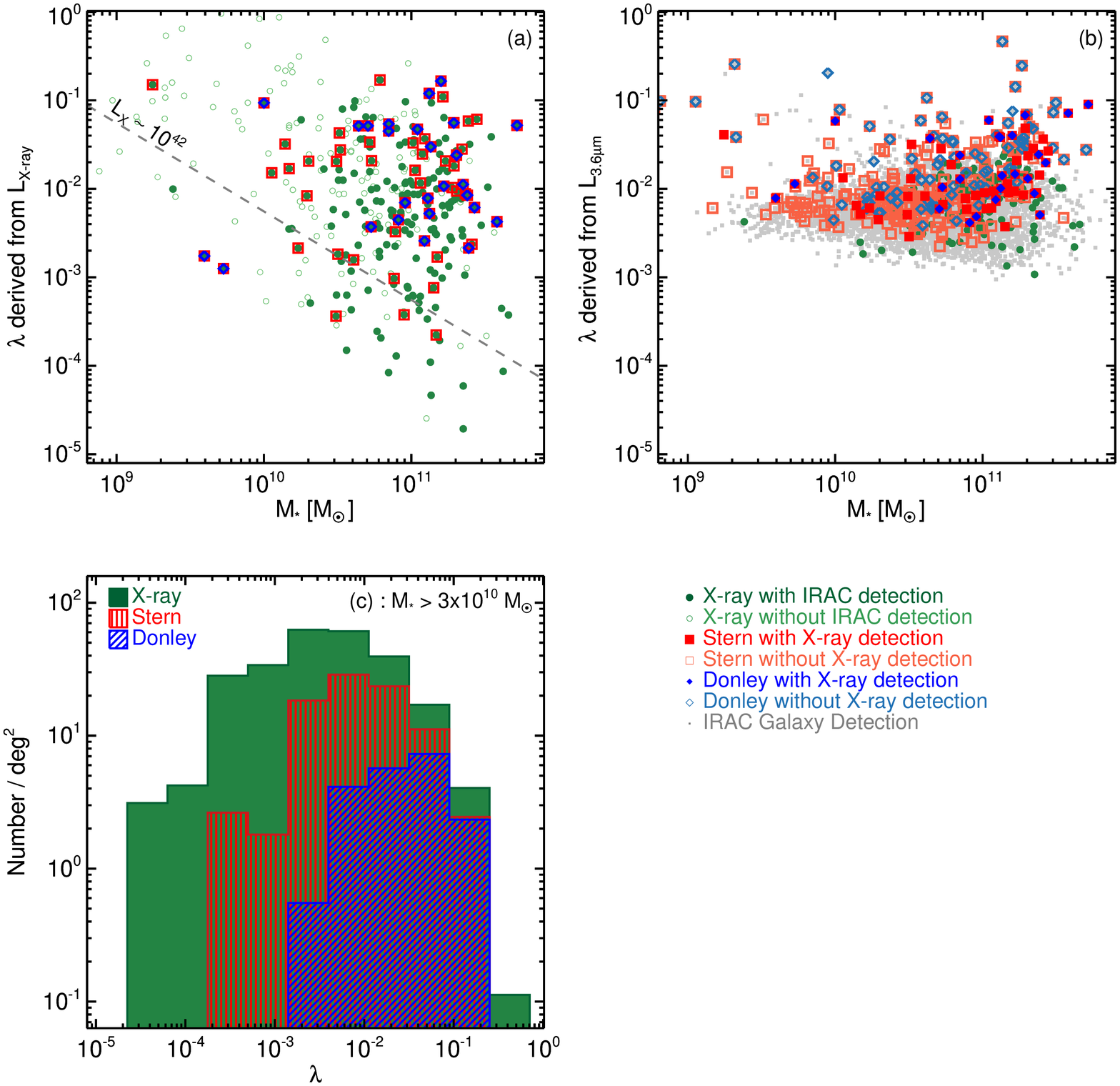}
  \caption{
(a) Specific accretion rate (\specific) derived from the X-ray luminosity vs. stellar mass.
For sources that are detected in both IRAC and X-ray (filled dark green circles) we also show \Stern or \Donley IR-AGN sources as red squares or blue diamonds, respectively.
The gray dashed line shows the approximate location of an \Lx{\sim}{42} source.
We find two populations of X-ray sources that are not identified as IR-AGNs: those that are not detected by IRAC and those that are detected by IRAC but do not meet either the \Stern or \Donley criteria.
Many X-ray AGNs that are not IRAC detected have high specific accretion rates (\Specific{>}{-1}) and have low stellar masses.
The sources that are IRAC detected but not identified by either selection technique are low-luminosity sources that have low specific accretion rates and high stellar masses.
(b) Specific accretion rate (\specific) derived from the \LIR\ luminosity vs. stellar mass.
X-ray detected sources are filled green circles, with filled red squares or blue diamonds for \Stern and \Donley IR-AGN, respectively.
\Stern and \Donley sources that are not X-ray detected are shown with open light red squares and light blue diamonds, respectively.
There is a floor near $10^{-2} < \specific < 10^{-3}$ where galaxy light begins to dominate the MIR emission.
(c) Specific accretion rate (\specific) distribution for the X-ray, \Stern and \Donley AGN samples for \mass{> 3 \times}{10}.
The \Stern and \Donley IR-AGN techniques tend to identify higher specific accretion rate sources relative to the X-ray AGN sample.
We include the X-ray completeness corrections for the X-ray detected sources.
}
\label{fig:eddington}
\end{figure*}

In this section, we investigate the optical rest-frame colors and stellar masses of the host galaxies for our IR-AGN and X-ray AGN samples with PRIMUS redshifts.
For broad-line AGNs, the optical light is dominated by the AGN rather than the host galaxy light; this prevents an accurate estimate of the host galaxy properties.
Therefore, in this section we exclude sources classified as broad-line AGNs, where the PRIMUS spectrum is better fit with $\chi^2$ of at least 50 by a broad-line AGN template compared to any galaxy template.
These broad-line AGNs are a sizable fraction of the X-ray AGN (\nXrayBroadLineFraction), \Stern IR-AGN (\nSternBroadLineFraction) and \Donley IR-AGN (\nDonleyBroadLineFraction) samples.

In the top panels of Figure~\ref{fig:colormass}, we plot the rest-frame optical $(u-g)$ colors and stellar masses for the X-ray AGN (left), \Stern IR-AGN (center), and \Donley IR-AGN (right column) samples.
We show the relevant AGN sample with black squares and PRIMUS galaxies with gray contours (30\%, 50\%, and  80\% contours) and gray points for individual galaxies outside of the 80\% contour.
In the left panel, we show the X-ray AGN that are selected by either of the \Stern IR-AGN or \Donley IR-AGN selection techniques with green diamonds.
In the center and right panels, we show IR-AGN that are also X-ray detected in green diamonds.
The stellar masses are estimated using the SED fitting code iSEDfit \citep[see ][for complete details]{Moustakas13}.
iSEDfit is a Bayesian fitting program that compares broadband photometry against large Monte Carlo grids of SED models that span a wide range of parameters (e.g. ages, metallicities, star formation histories, star bursts, and dust content).
We construct our stellar mass models using the \citet{Bruzual03} stellar population synthesis models assuming the \citep{Chabrier03} initial mass function from 0.1 to 100$\mstar$.
We assume a uniform stellar metallicity prior in the range $0.004<Z<0.04$, and smooth, exponentially declining star formation histories, $\psi(t) \propto e^{-\gamma t}$, with $\gamma$ drawn uniformly from the interval [0.01, 1] Gyr$^{-1}$, and we include bursts of star formation.
In the bottom panels of Figure~\ref{fig:colormass}, we calculate the fraction of all PRIMUS galaxies that are detected in one of the AGN selection methods as a function of stellar mass for red and blue sources.
We define each source as blue or red using the magnitude and redshift dependent cut from \citet{Aird12}
\begin{equation}
  C = (u-g) - (0.671 - 0.031{M_g} - 0.065{z}),
\end{equation}
where $M_g$ is the absolute $g$-band magnitude and $z$ is the redshift. 
To calculate the number density and account for the variation in survey depth, we use the X-ray completeness corrections (X-ray sensitive area for a given X-ray flux) for the sources that are X-ray detected and the IR--optical overlap area for all other sources.

We find that the fraction of galaxies with detected X-ray AGNs increases with stellar mass.
The fraction of blue galaxies with X-ray AGNs is a factor $\sim$2 higher than for red galaxies (across all stellar masses).
These trends are in agreement with \citet{Aird12}, where the increase with stellar mass is attributed to a selection effect. 
AGNs in massive galaxies appear more luminous (for the same specific accretion rate scaled relative to the host stellar mass) and so are easier to detect.
AGNs selected by either \Stern IR-AGN or \Donley IR-AGN techniques follow the same basic trend, although both techniques generally identify a lower fraction of AGNs in both red and blue galaxies at all stellar masses.
The rate that the fraction of red galaxies with \Stern or \Donley AGNs increases with stellar mass is much slower than for red galaxies with X-ray AGNs.
This lower rate of change for the IR-AGN selections suggests that they are less efficient at identifying AGNs in red, massive galaxies (\mass{\gtrsim}{10.5}).
The fraction of galaxies with \Stern IR-AGNs is \emph{higher} than the X-ray fraction at low masses which may be due to contamination of \Stern IR-AGNs by star-forming galaxies.

In Figure~\ref{fig:eddington}(a), we estimate the specific accretion rate (\specific) of the X-ray-detected sample as a function of the host stellar mass.
Following \citet{Aird12}, we define the specific accretion rate from the bolometric luminosity derived from the X-ray luminosity using the \citet{Hopkins07} quasar bolometric corrections\footnote{\url{http://www.tapir.caltech.edu/~phopkins/Site/qlf}}.
For the hard X-ray-selected sample, this is dominated by the emission from the AGN with possibly only minor contribution from the host galaxy.
From the bolometric luminosity, we calculate the specific accretion rate
\begin{equation}
\lambda = 
\frac{L_\mathrm{bol}}{L_\mathrm{Edd}} = 
\frac{L_\mathrm{bol}}{1.3\times10^{38}\;\ergs \times 0.002 \dfrac{\mstar}{\msun}},
\end{equation}
\noindent where \mstar\ is the host galaxy stellar mass and $L_\mathrm{bol}$ is the bolometric luminosity.
The X-ray sources with IRAC detections are solid green circles and the sources without IRAC detections are open green circles.
Sources with IRAC detections that are selected by either \Stern or \Donley IR-AGN selection technique are outlined with red squares or blue diamonds.
The vast majority of \Donley IR-AGNs also satisfy the \Stern IR-AGN selection criteria.
This figure shows two populations of X-ray sources that are not identified as IR-AGNs: those that are not detected by IRAC (see Section~\ref{sec:depth}) and those that are detected by IRAC but do not meet either the \Stern or \Donley criteria.
The former populations consists primarily of sources at low stellar masses and high specific accretion rates.
These sources are primarily found in the fields that have shallower IRAC data compared to their X-ray data (\ESONE, \XMM, and \COSMOSA).

The majority of the X-ray AGNs that are not \Stern or \Donley selected (but do have IRAC detections: green solid symbols in Figure~\ref{fig:eddington}(a)) have large stellar masses (\mass{> 3\times}{10}), low-to-moderate X-ray luminosities (\Lx{<}{43}), and low specific accretion rates (\SpecificP{<}{1}).
Additionally, many of these sources have a moderate-to-large MIR to X-ray luminosity ratios (\Lratio $>$ 1).
In fact, their \Lratio\ ratios are larger than typically found for the IR-AGN population and for optically selected quasars \citep[e.g.][]{Richards06}.
This suggests that stellar light from their massive host galaxies is dominating the emission in the IRAC bands.
Indeed, for a stellar mass-to-light ratio of order unity \citep[e.g.][]{Bell01} and \Lratio\ for a pure AGN SED \citep{Richards06}, a galaxy with a specific accretion rate of \SpecificP{=}{1}\ would have comparable emission at 3.6~\um\ from stellar light and from the central AGN.
At lower specific accretion rates, the host galaxy would then dominate the emission in the IRAC bands.
This provides a natural explanation for why many X-ray AGNs at lower specific accretion rates are not selected as IR-AGNs.
These findings are in good agreement with \citet{Donley12}, who show that the average SED of luminous AGNs is missed due to a noticeable 1.6~\um\ stellar bump that prevalent in older, massive galaxies.

In Figure~\ref{fig:eddington}(b), we estimate the specific accretion rate based on MIR luminosity.
The specific accretion rates for these sources are less well constrained due to the possible addition of galaxy light to the observed MIR flux.
To ensure that the bolometric luminosity distribution are similar for the X-ray and IRAC overlapping sample, we first use the \citet{Richards06} mean quasar SED to estimate the \LX\ luminosity from the \LIR\ luminosity (see the orange line in Figure~\ref{fig:excess}).
We apply the \citet{Hopkins07} hard X-ray bolometric correction to the equivalent X-ray luminosity to calculate the specific accretion rate for these sources.
We do not use the  \citet{Hopkins07} bolometric corrections for the IR luminosity due to the overprediction of the bolometric luminosity for IR-AGN sources that are also X-ray detected.
This method for estimating specific accretion rates illustrates the floor near $10^{-2} < \specific < 10^{-3}$, where galaxy light dominates the MIR luminosity.

In Figure~\ref{fig:eddington}(c), we show the specific accretion rate distribution for \Stern IR-AGNs, and \Donley IR-AGNs and X-ray AGNs.
We restrict this comparison to AGNs in host galaxies with stellar mass \mass{> 3\times}{10}, due to incompleteness at lower stellar masses.
For all AGNs with an X-ray detection, we use the specific accretion rate derived from the X-ray luminosity.
For the IR-AGNs that are not detected in the X-rays we use their estimated specific accretion rate from their MIR luminosity.
Some of the IR-AGNs with X-ray detections have significant absorption in the X-rays (e.g., Figure~\ref{fig:excess}).
For IR-AGNs with \Lratio $>$ 10, we use the specific accretion rates derived from their
MIR luminosities.
We note that this does not have a significant effect on the lambda distributions in Figure~\ref{fig:eddington}(c).
Both the \Stern and \Donley IR-AGN samples have specific accretion rate distributions that do not match the X-ray distribution at the $> 99.7\%$ confidence level from using a two-sided weighted KS test.
In particular, the X-ray sample includes a significant number of sources at low to moderate specific accretion rates that are not identified as IR-AGNs.
These low to moderate specific accretion rate sources are preferentially found in red, high-mass galaxies, which explains the larger fraction of red host galaxies identified with X-ray AGNs compared to IR-AGN selection techniques.


\section{Discussion}\label{sec:discussion}
In this paper we compare AGN samples selected in the MIR to those selected using hard-band X-rays, in order to quantify both the overlap and uniqueness of each selection method as well as understand the AGN populations identified.
We use the \Stern and \Donley IR-AGN selection methods (described in Section~\ref{sec:sample}) to identify IR-AGN samples using \spitzer/IRAC data in four fields that have X-ray coverage from \chandra\ or \xmm:  CDFS, COSMOS, ES1, and the XMM field.
We create IR-AGN and X-ray AGN samples to various depths in both the IR and X-ray data and find that the number density of AGNs recovered with each selection method varies strongly as a function of depth, as does the overlap between the IR-AGN and X-ray AGN samples.
We use PRIMUS spectroscopic redshifts in these fields to study the AGN luminosities, host galaxy colors, and host stellar masses of the IR-AGN and X-ray AGN samples and investigate potential contamination in IR-AGN selection.

There are several advantages to our approach, compared to previous studies.
We use large AGN samples (a total of \nTotalStern\ \Stern IR-AGNs, \nTotalDonley\ \Donley IR-AGNs and \nTotalXray\ X-ray AGNs) spanning multiple fields, which allows us to perform statistical comparisons as a function of joint parameters, while minimizing the effects of cosmic variance.
We take advantage of the large number of spectroscopic redshifts in these fields provided by PRIMUS 
(a total of \nPrimusRedshifts\ AGNs with PRIMUS redshifts between 0.2$ < z < $1.2), 
so that we do not have to rely on photometric redshifts which have larger errors.
We further study the overlap and uniqueness of IR-selected versus X-ray-selected AGNs as a function of both the depth of the IR data and the depth of the X-ray data.
With a large dynamic range in survey depth, we can test how the completeness, purity, and contamination of the AGN selection depends on depth.
Such tests are required to reconcile the often vastly different and seemingly conflicting results in the literature regarding IR and X-ray AGN selection.

Another key difference in our methodology is that we use hard-band (2--10 keV) X-ray selection, as opposed to soft or full band, and we use weights for X-ray detections to account for incompleteness in the X-ray data.
Hard-band X-ray selection is sensitive to both unabsorbed and moderately absorbed AGNs, though it will miss heavily absorbed, Compton-thick AGNs.
Soft (0.5--2 keV) or full (0.5--10 keV) band selection, by comparison, is biased toward unabsorbed sources (with column densities of \Nh{<}{22}).
We account for the large positional uncertainties in the X-ray data by using the likelihood ratio matching technique to securely match the X-ray sample to both the IR and optical samples. 
Finally, we account for the varying sensitivity of the X-ray data across individual fields by applying statistical completeness corrections to all X-ray sources.
This last correction is crucial to accurately quantify the number density of X-ray AGNs as a function of depth and robustly compare with IR-AGN samples.
We emphasize that this correction is not generally applied in the literature, where studies simply compare the fraction of IR-AGNs that are observed in X-ray data, without taking into account incompleteness in the X-ray data.

\subsection{How AGN Selection Varies with IR and X-Ray Depth}\label{sec:discussagn}
We first consider the fraction of IR-AGNs that are detected in X-rays, as a function of both the IR and X-ray data depth.
Generally, a higher fraction of \Donley IR-AGNs are X-ray detected than in the \Stern IR-AGN sample, consistent with our findings above that the \Stern IR-AGN selection can suffer from contamination.  
At the relatively shallow IR depth of the \swire\ survey, 60\% of IR-AGN are X-ray detected when the X-ray depth is comparable to the IR depth.
Using the deepest IR and X-ray data, the fraction of IR-AGNs that are X-ray detected rises to $\sim$70\% -- $\sim$80\%.
However, this fraction rises to $\sim$90\% when combining shallow IR and deep X-ray data, indicating that the vast majority of IR-AGNs are indeed X-ray emitting AGNs.  

As discussed above, IR-AGN selection methods identify additional AGNs not detected in X-rays, when the IR and X-ray data are of comparable depth.
At shallow IR depths the \Stern and \Donley techniques recover very similar number densities of AGNs with a large overlap between the populations.
As IR depth increases, both the \Stern and \Donley selections identify a larger population of AGNs.
However, the number density recovered by the \Stern selection increases at a greater rate --- in our deepest IR data \Stern recovers approximately twice the number density of AGNs as the \Donley IR-AGN selection.
As discussed in Section~\ref{sec:contamination}, much of the relative increase is likely due to contamination.
For any depth IR and X-ray data, the addition of \Stern IR-AGN sample to X-ray AGN sample increases the total AGN sample by 7\%$-$27\%, 
depending on the depth of the IR and X-ray data. 
In comparison, addition of \Donley IR-AGN selection to X-ray AGN selection increases the total AGN sample by 4\%$-$20\%.
Both IR-AGN selection methods increase the total AGN samples more at shallower IR and X-ray depths.
At all IR and X-ray depths studied here, X-ray AGN selection identifies a higher number density of AGNs than either IR-AGN selection method.

It is also extremely important to account for the well known and easily characterized incompleteness of X-ray selection.
The variation in X-ray sensitivity over a field can substantially reduce the number of IR-AGNs with observed X-ray detections.
It is vital to account for this effect to understand any true, underlying differences between AGN populations.

At all IR and X-ray depths studied here we find a much higher fraction of IR-AGNs are detected in X-rays, compared to values in the literature.
This is due to the X-ray completeness corrections that we make, to account for the varying X-ray sensitivity within a field.
If we do not apply these completeness weights, we find similar fractions to those in the literature.
For example, in the XBo\"{o}tes survey with shallow IR (\fir{\sim}{51}) and shallow X-ray (\fx{\sim}{-15}) data, \citet{Hickox09} find that 38\% of their \Stern IR-AGNs are X-ray detected.
Using similar depth data if we do not apply X-ray completeness corrections we find that 37\% of \Stern IR-AGNs are X-ray detected; however this fraction rises to 63\% after applying the completeness corrections.
\citet{Donley07} use deep IR (\fir{\sim}{14.5}) and medium-depth X-ray (\fx{\sim}{-15}) data in the CDF-North field to find that 33\% of the \Stern IR-AGN are X-ray detected.
Using similar depth data if we do not apply X-ray completeness corrections we find a similar fraction of 35\% of \Stern IR-AGNs are X-ray detected; however this fraction rises to 57\% after applying the completeness corrections. 
\citet{Park10} use very deep IR data (\fir{\sim}{6.3}) and medium-depth X-ray (\fx{\sim}{-16}) data in the EGS field and find that a comparatively low fraction of \Stern IR-AGNs are X-ray detected.
As their X-ray data are not as deep as their IR data, it is not surprising that they find a low fraction.
We do not have comparably deep IR data, however at a similar X-ray depth using the deepest IR data we have, we find that at most 35\% of IR-AGNs should be X-ray detected, if X-ray completeness is not accounted for. 
This upper limit is consistent with their value.

Thus, we find that the wide range of X-ray detected IR-AGN fractions reported in the literature can be accounted for by the varying depths of both the IR and X-ray data used in these studies.
We also find that accounting for X-ray incompleteness, which is generally not done in the literature, increases this fraction by a factor $\sim$2 which results in 25\% more of the IR-AGNs being counted (statistically) as X-ray detected.
This illustrates how differing depths of data can lead to very different conclusions in prior studies of IR and X-ray AGNs.

\subsection{Contamination and Bias of IR-AGN Selection}\label{sec:discusscontamination}
Much, if not most, of the difference between the \Stern IR-AGN sample and \Donley IR-AGN sample is due to contamination by galaxies that do not host an AGN (or at least, are not dominated by a luminous AGN in the MIR).
Using PRIMUS redshifts, we study the evolution of the distribution of \Stern and \Donley IR-AGNs and X-ray-detected AGNs in IRAC color--color space from $z$ $\sim$ 0.2 to $z$ $\sim$ 1.2 and compare our results to the predicted evolution of star-forming and quiescent galaxy and AGN SED templates in this space.
We find that the \Stern IR-AGNs sample is contaminated at $z$ $\sim$ 0.3 by star-forming galaxies and at $z$ $\sim$ 1.1 by quiescent galaxies.
This is reflected in the fraction of \Stern IR-AGN that are not detected in X-rays; for the full \Stern IR-AGN sample with PRIMUS redshifts, spanning $0.2 < z < 1.2$, this fraction is \nSternAllXrayMissingFraction, while at $0.2 < z < 0.45$ it is \nSternLowXrayMissingFraction\ and at $1.0 < z < 1.2$ it is \nSternHighXrayMissingFraction\ when using the X-ray completeness weights.

Extending our analysis of the galaxy and AGN templates to higher redshift, beyond the reach of PRIMUS, we find that the \Stern IR-AGN samples will be contaminated by high-redshift star-forming galaxies.
We do not find any galaxy contamination in the \Donley IR-AGN sample; this is reflected by the lack of evolution in the fraction of \Donley IR-AGNs that are not X-ray detected.
Using the \Donley IR-AGN sample as a baseline, we find that galaxy contamination in the \Stern IR-AGN selection is significant only for IR surveys deeper than the SWIRE limit of \fir{\sim}{100}.

Overall, we find that the \Donley IR-AGN selection is less complete than the \Stern IR-AGN selection.
This is demonstrated by the fact that a smaller fraction of X-ray AGNs are also \Donley IR-AGNs.
Additionally there are correctly identified AGNs within the \Stern IR-AGN sample that are not found by the \Donley IR-AGN selection.
However, the \Donley IR-AGN selection is more reliable than the \Stern IR-AGN selection, in that a higher fraction of \Donley IR-AGNs are also X-ray detected, and the \Donley IR-AGN selection does not suffer from galaxy contamination.
The \Stern and \Donley IR-AGN samples demonstrates the importance of finding a balance between minimizing both contamination and incompleteness.

Both the \Donley and \Stern IR-AGN selections appear to be biased, however, in that they select high luminosity AGN.
This is seen in the X-ray luminosity distributions of the IR-AGN samples compared to the X-ray AGN sample, where the IR-AGN selection techniques preferentially sample the population of \Lx{\gtrsim}{43.5} AGN.
These high luminosity sources have IRAC colors similar to power-law AGNs and are easily detected with shallow IR surveys.

\subsection{Uniqueness of IR-AGN Selection}\label{sec:discussuniq}
Even with a relatively deep X-ray survey and shallow IR data, we find that there is at least $\sim$10\% of the IR-AGN population which is not detected in X-rays even after correcting for the variable X-ray sensitivity.
For samples of comparable X-ray and IR depth, this fraction is typically $\sim$20\% -- $\sim$30\%. 
This reflects the \textit{uniqueness} of IR-AGN selection, in that these are sources that are only identified in the IR and would typically be missed in X-ray surveys.  
What kind of AGNs does IR selection uniquely identify?
These AGN could in theory be missed by X-ray selection either because they are intrinsically less X-ray luminous, due to a lower accretion rate onto the supermassive black hole, or they could be obscured by a high-column density of gas and dust, limiting the ability of even deep X-ray surveys to detect them.
For the population of IR-AGNs that are X-ray detected, as discussed above we find that IR-AGN selection techniques preferentially select a population of luminous AGNs with \Lx{\gtrsim}{43.5}.
However, there is a small fraction ($\sim$8\%) of IR-AGN with very high IR to X-ray luminosity ratios, implying heavy obscuration.  
Therefore, up to $\sim$20\% of IR-AGNs could be sources with moderate to heavy obscuration.  

In comparing the HR distributions of IR- and X-ray-selected AGNs, we find no statistically significantly difference in the distributions for X-ray AGNs with higher or lower X-ray luminosity.
This comparison does not account for the $\sim$10\% of IR-AGN sources that are not X-ray detected.
Thus, while the majority of IR-AGNs are also X-ray AGNs, the IR selection techniques adds a small but important population of obscured sources that are missed even with the deepest X-ray surveys.
Additionally, the 10\% of IR-AGN sources that are not X-ray detected may be Compton-thick and thus consistent with X-ray background synthesis models \citep[e.g.,][]{Gilli07}.

\subsection{Host Galaxies}
With the exception of broad-line sources, the optical SEDs of the IR-AGN and X-ray AGN samples are dominated by light from the host galaxy, which allows us to determine galaxy properties for these sources.
Generally, the \Stern and \Donley IR-AGN samples follow the same trends as the X-ray population, showing that IR-AGNs and X-ray AGNs do \textit{not} populate vastly different host galaxy populations.
The galaxy hosts span a similar range of stellar masses (\mrange{10}{11.5}) and optical colors compared to the PRIMUS parent galaxy sample.
The \Donley IR-AGN and X-ray AGN samples identify similar fractions of AGNs in blue, star-forming hosts and red, quiescent hosts.
These fractions of red and blue AGN hosts are similar to the red and blue fractions for all PRIMUS galaxies.

\citet{Hickox09} find that X-ray AGNs are preferentially found in galaxy hosts in the green valley, the minimum of the optical color bimodality between the red sequence and the blue cloud \citep{Martin07}, whereas \Stern IR-AGNs are typically in blue host galaxies.
We do not agree with these results, which did not correct for X-ray incompleteness and contamination in the \Stern sample.
Most recent papers \citep[e.g.,][]{Silverman09, Xue10, Aird12} find that the X-ray AGN fraction is highly stellar mass dependent.
Similar to \citet{Nandra07}, \citet{Silverman09} and \citet{Aird12} we find that at a given stellar mass the fraction of galaxies with X-ray AGN is higher in the blue cloud.
In particular for the green valley \citet{Nandra07} using the $\sim$200 ks X-ray data in the Extended Groth Strip to find that 13\% of X-ray-detected AGNs have green host galaxies, which is consistent with our X-ray AGN sample (9\%) using a similar green valley definition.
If X-ray completeness correction weights are not applied, the fraction of green host galaxies with X-ray AGNs rises only slightly, from 9\% to 11\%.

For the \Stern IR-AGNs, we generally find equal fractions of sources with blue or red host galaxies.
There is a slightly larger fraction with blue hosts when compared using stellar-mass matched samples, but this is mainly due to blue star-forming host galaxies contaminating the sample rather than differences in the underlying population.
The \Donley IR-AGN sample is more evenly distributed between red and blue host galaxies at fixed stellar mass.

We do find that the fraction of galaxies hosting an AGN increases with galaxy stellar mass. 
The fraction of massive (\mass{> }{11}) galaxy hosts is larger for the X-ray AGN sample (\nXrayMassiveFraction) compared to the \Stern IR-AGN (\nSternMassiveFraction) or \Donley IR-AGN sample (\nDonleyMassiveFraction).
Conversely, the fraction of less massive (\mrange{9}{10}) galaxy hosts for the X-ray AGN sample (\nXrayMediumFraction) is smaller than either the \Stern IR-AGN sample (\nSternMediumFraction) or the \Donley IR-AGN sample (\nDonleyMediumFraction). 
However, these differences are minor compared to the similarity of the host galaxy stellar mass distribution between any of the AGN selection techniques.

From the host galaxy stellar masses and the estimated AGN bolometric luminosities we infer the specific accretion rates, which provide a valuable measure of the intrinsic differences of the selected AGN populations.
While IR-AGN samples are generally found in similar host galaxies as X-ray AGNs, we find that IR-AGNs have relatively high specific accretion rates.
X-ray AGNs that are not identified using IR-AGN techniques include both sources that are not detected by IRAC and have high accretion rates and low stellar masses and sources that are detected by IRAC that have low to moderate specific accretion rates and high stellar masses.
Some of the sources that are not IRAC detected are from fields (\ESONE, \XMM, and \COSMOSA) with shallow IRAC data.
For these luminous sources, we do not find any significant difference in the HRs of the X-ray AGN and IR-AGN samples, suggesting similar numbers of obscured sources.
The X-ray sources with IRAC detections that are not IR-AGN selected have lower specific accretion rates and reside in red, massive galaxies and have large IR to X-ray luminosities.
This suggests that they are not being identified by the IR-AGN techniques due to a bright galaxy component in the MIR due to the 1.6~\um\ stellar bump which is dominating the AGN light.
They contribute to the large rise in the fraction of red massive galaxies that have a X-ray AGN.
This accounts in part for the significant population of X-ray AGNs that is not identified by either IR-AGN technique.


\section{Conclusions}\label{sec:conclusions}
In this paper, we compare the completeness, contamination, overlap and uniqueness of \spitzer/IRAC and X-ray-identified AGNs.
We quantify the differences in X-ray AGNs and IR-AGN selection techniques due to IR and X-ray survey depths, X-ray obscuration, and survey completeness to assess the usefulness of these techniques to identify obscured AGNs.
We use \spitzer/IRAC data, \xmm, and \chandra\ X-ray data at multiple depths to construct the largest sample of IR and X-ray-selected AGNs to date.
We focus on the \citet{Stern05} and \citet{Donley12} selection techniques which select AGNs that dominate the MIR with a red power-law SED.
We statistically compare the IR-AGN populations against X-ray-selected AGN samples.
The combination of multiple X-ray and IRAC depth surveys in four fields (CDFS, ES1, COSMOS, and XMM) allows us to study the overlap and uniqueness of the AGN selections as a function of both IR and X-ray depth; having multiple fields also reduces cosmic variance.
This gives us a large sample with \nTotalStern\ \Stern IR-AGNs, \nTotalDonley\ \Donley IR-AGNs and \nTotalXray\ X-ray AGNs.
Characterizing the variation of the samples as a function of survey depth also allows us to probe the bias and contamination of individual identification techniques.

We take advantage of the $\sim$1,500 secure PRIMUS redshifts in these fields to probe the intrinsic properties of the AGNs, including luminosity, specific accretion rate, and HR, as well as properties of the host galaxies, including color and stellar mass.
We compare the AGN properties of the IR-AGN versus X-ray AGN samples and compare their host galaxies to the full PRIMUS galaxy population.
The main results from our work are as follows:

\begin{enumerate}
\item IR-AGN selection identifies predominantly luminous AGNs, with \Lir{~\sim~}{44.5} and \Lx{\sim}{43.5}.
X-ray AGN selection identifies a larger population of AGNs, including those with lower luminosities and/or lower accretion rates, where the host galaxy light dominates the MIR emission.
These AGNs are found throughout IRAC color--color space, with the bulk not easily identified using IR-AGN selection.
These results indicate that IR-AGN selection techniques are not as efficient as X-ray selection in identifying complete AGN samples.

\item IR-AGN selection does not identify a substantial population of obscured AGNs relative to X-ray samples.
However, $\sim$10\% of IR-AGNs are not detected in extremely deep X-ray data, setting an upper limit on the fraction that could be very heavily obscured, Compton-thick sources.

\item \Stern IR-AGN selection is contaminated by non-AGNs at specific redshifts ($z$ $\sim$ 0.3, $z$ $\sim$ 1.1, and $z$ $\gtrsim$ 2.5).
The level of contamination depends on depth and is not significant at shallow (e.g., \swire) IR depths.
\Donley IR-AGN selection is not contaminated.

\item IR-AGN and X-ray AGN samples both preferentially identify AGNs in massive galaxies.
They further both identify AGNs in red and blue host galaxies, with a similar ratio of red to blue galaxies.
The host galaxy stellar masses and colors are therefore quite comparable between IR and X-ray AGN selection.

\item Both the \Stern and \Donley IR-AGN samples identify AGNs with high specific accretion rates relative to the X-ray-detected AGN sample.
The low and moderate accretion rate AGNs identified only in X-rays have large stellar mass host galaxies, which dominate the MIR SED.
There are also high accretion rate X-ray AGNs that are not identified by either IR-AGN technique, which lack IRAC detections.
\end{enumerate}

We find that the majority of current X-ray and IR surveys in cosmological fields are often not well matched in depth, in that the X-ray data are generally significantly deeper than the IR data.
For example, a pure AGN SED shows that \goods-depth IRAC data and $\sim$20 ks X-ray data will probe roughly the same intrinsic AGN luminosity.
Therefore, while existing deep X-ray surveys ensure that we have a fairly complete X-ray AGN sample, deeper IR surveys are needed to build a comparably complete IR AGN sample.
Moreover, larger shallow and wide X-ray surveys can be used to accurately compare to the luminous IR-AGN samples from these shallower and wider IR surveys.

IR and X-ray AGN selections appear to identify fairly similar AGN populations in similar high stellar mass galaxies, with no strong preference for either red or blue host galaxies.
There is a large overlap between these selections, though IR selection preferentially identifies brighter AGNs.
However IR-AGN selection techniques do identify a small population ($\sim$10\%) that is not identified in extremely deep X-ray surveys and could be very heavily obscured.
A combined IR and X-ray AGN selection will identify a more complete sample than either selection alone, including both heavily obscured AGNs and AGNs that are under-luminous relative to their host galaxies.

\vspace{2em}

We thank our anonymous referee and Jennifer Donley for useful comments that have improved this paper.
We gratefully acknowledge helpful discussions and feedback from Ramin Skibba.
Funding for PRIMUS has been provided by NSF Grants AST-0607701, 0908246, 0908442, 0908354, and NASA Grant 08-ADP08-0019.
ALC acknowledges support from the Alfred P. Sloan Foundation and NSF CAREER Award AST-1055081.
AJM and JA acknowledge support from NASA Grant NNX12AE23G through the Astrophysics Data Analysis Program.

We thank the CFHTLS, COSMOS, DLS, and SWIRE teams for their public data releases and/or access to early releases.
This paper includes data gathered with the 6.5 m Magellan Telescopes located at Las Campanas Observatory, Chile.
We thank the support staff at LCO for their help during our observations, and we acknowledge the use of community access through NOAO observing time.
Some of the data used for this project are from the CFHTLS public data release, which includes observations obtained with MegaPrime/MegaCam, a joint project of CFHT and CEA/DAPNIA, at the Canada-France-Hawaii Telescope (CFHT) which is operated by the National Research Council (NRC) of Canada, the Institut National des Science de l'Univers of the Centre National de la Recherche Scientifique (CNRS) of France, and the University of Hawaii. This work is based in part on data products produced at TERAPIX and the Canadian Astronomy Data Centre as part of the Canada-France-Hawaii Telescope Legacy Survey, a collaborative project of NRC and CNRS.
We also thank those who have built and operate the Chandra and XMM-Newton X-Ray Observatories.
This research has made use of the NASA/IPAC Infrared Science Archive, which is operated by the Jet Propulsion Laboratory, California Institute of Technology, under contract with the National Aeronautics and Space Administration.

\begin{deluxetable}{lrrrr}  
  \tablecolumns{5}
  \tablewidth{0pt}
  \tablecaption{IRAC and X-Ray Flux Limits. \label{table:fluxlimit}}
  \tablehead{Field & \colhead{X-Ray Depth} & \colhead{IR Depth} & \colhead{\makecell[c]{IRAC $f_{5.8\um}$\\ Limit [\uJy]}} &    \colhead{\makecell[c]{X-ray $f_\mathrm{X}$\\ Limit [\ergscm]}}  }
  \startdata
              CDFS &             $\sim$2Ms &             \goods &                                                     20.0 &                                        1.8$\times10^{-16}$ \\
                   &                       &            \cosmos &                                                     44.7 &                                        1.8$\times10^{-16}$ \\
                   &                       &             \swire &                                                    100.0 &                                        1.8$\times10^{-16}$ \\
                   &           $\sim$200ks &             \goods &                                                     20.0 &                                        7.1$\times10^{-16}$ \\
                   &                       &            \cosmos &                                                     44.7 &                                        7.1$\times10^{-16}$ \\
                   &                       &             \swire &                                                    100.0 &                                        7.1$\times10^{-16}$ \\
  \hline
            COSMOS &           $\sim$160ks &            \cosmos &                                                     44.7 &                                        7.1$\times10^{-16}$ \\
                   &            $\sim$40ks &                    &                                                     44.7 &                                        2.8$\times10^{-15}$ \\
  \hline
               ES1 &            $\sim$40ks &             \swire &                                                    100.0 &                                        2.8$\times10^{-15}$ \\
  \hline
               XMM &        $\sim$50-100ks &             \swire &                                                    100.0 &                                        7.1$\times10^{-16}$ \\
                   &         $\sim$10-50ks &                    &                                                    100.0 &                                        5.6$\times10^{-15}$   
  \enddata
\end{deluxetable}

\begin{deluxetable*}{lrrrrrrrrr}  
  \tablecolumns{10}
  \tablewidth{0pt}
  \tablecaption{Field Areas and Overlap \label{table:area}}
  \tablehead{ &                       \multicolumn{2}{c}{} &                \multicolumn{3}{c}{Area [deg$^{2}$]} &                                                                                                                        \multicolumn{4}{c}{Overlap Area [deg$^{2}$]} \\
  Field       & \colhead{X-Ray Depth} & \colhead{IR Depth} & \colhead{X-ray} & \colhead{IRAC} & \colhead{PRIMUS} & \colhead{\makecell[c]{IRAC\\X-Ray}} & \colhead{\makecell[r]{PRIMUS\\IRAC}} & \colhead{\makecell[r]{PRIMUS\\X-Ray}} &    \colhead{\makecell[r]{PRIMUS\\IRAC\\X-Ray}}  }
  \startdata
         CDFS &             $\sim$2Ms &             \goods &            0.12 &           0.06 &             2.90 &                                0.06 &                                 0.06 &                                  0.12 &                                           0.06 \\
              &                       &            \cosmos &            0.12 &           0.44 &             2.90 &                                0.12 &                                 0.21 &                                  0.12 &                                           0.12 \\
              &                       &             \swire &            0.12 &           7.24 &             2.90 &                                0.12 &                                 2.57 &                                  0.12 &                                           0.12 \\
              &           $\sim$200ks &             \goods &            0.31 &           0.06 &             2.90 &                                0.06 &                                 0.06 &                                  0.18 &                                           0.06 \\
              &                       &            \cosmos &            0.31 &           0.44 &             2.90 &                                0.30 &                                 0.21 &                                  0.18 &                                           0.18 \\
              &                       &             \swire &            0.31 &           7.24 &             2.90 &                                0.31 &                                 2.57 &                                  0.18 &                                           0.18 \\
  \hline
       COSMOS &           $\sim$160ks &            \cosmos &            1.00 &           2.61 &             1.18 &                                1.00 &                                 1.18 &                                  0.81 &                                           0.81 \\
              &            $\sim$40ks &                    &            2.16 &           2.61 &             1.18 &                                2.16 &                                 1.18 &                                  1.18 &                                           1.18 \\
  \hline
          ES1 &            $\sim$40ks &             \swire &            0.64 &           6.03 &             1.03 &                                0.64 &                                 1.03 &                                  0.59 &                                           0.59 \\
  \hline
          XMM &        $\sim$50-100ks &             \swire &            1.07 &           8.28 &             3.41 &                                1.03 &                                 3.37 &                                  0.91 &                                           0.88 \\
              &         $\sim$10-50ks &                    &            5.43 &           8.28 &             3.41 &                                4.07 &                                 3.37 &                                  1.30 &                                           1.30   
  \enddata
\end{deluxetable*}

\begin{deluxetable*}{lrrrrrrrrr}  
  \tablecolumns{10}
  \tablewidth{0pt}
  \tablecaption{Raw Numbers and Number Densities of IR-AGN and X-ray AGN samples in Each Field with Overlapping IRAC and X-Ray Data \label{table:numberweight}}
  \tablehead{ &                 \multicolumn{2}{c}{Limits} &            \multicolumn{3}{c}{Weighted Number/deg$^2$ (Raw Number)} &                                                                                                                                                        \multicolumn{4}{c}{Overlap Weighted Number/deg$^2$ (Raw Number)} \\
  Field       & \colhead{X-Ray Depth} & \colhead{IR Depth} & \colhead{X-Ray\tablenotemark{a}} & \colhead{Stern} & \colhead{Donley} & \colhead{\makecell[r]{Stern\\Donley}} & \colhead{\makecell[r]{Stern\\X-Ray\tablenotemark{a}}} & \colhead{\makecell[r]{Donley\\X-Ray\tablenotemark{a}}} &    \colhead{\makecell[r]{Stern\\Donley\\X-Ray\tablenotemark{a}}}  }
  \startdata
         CDFS &             $\sim$2Ms &             \goods &                       5055 (176) &      2477 (115) &        1118 (53) &                             1118 (53) &                                             1769 (74) &                                               928 (42) &                                                         928 (42) \\
              &                       &            \cosmos &                       4912 (245) &       1135 (91) &         709 (59) &                              709 (59) &                                              996 (74) &                                               627 (49) &                                                         627 (49) \\
              &                       &             \swire &                       4912 (245) &        336 (33) &         269 (26) &                              269 (26) &                                              303 (29) &                                               244 (23) &                                                         244 (23) \\
              &           $\sim$200ks &             \goods &                        2204 (81) &       1598 (84) &         729 (38) &                              729 (38) &                                              752 (35) &                                               436 (21) &                                                         436 (21) \\
              &                       &            \cosmos &                       2077 (418) &       961 (240) &        521 (139) &                             521 (139) &                                             787 (187) &                                              436 (113) &                                                        436 (113) \\
              &                       &             \swire &                       2073 (421) &        349 (96) &         244 (69) &                              244 (69) &                                              317 (86) &                                               218 (61) &                                                         218 (61) \\
  \hline
       COSMOS &           $\sim$160ks &            \cosmos &                      2067 (1176) &       972 (766) &        556 (443) &                             549 (436) &                                             717 (510) &                                              438 (324) &                                                        433 (320) \\
              &            $\sim$40ks &                    &                       750 (1104) &      700 (1279) &        408 (744) &                             403 (735) &                                             399 (628) &                                              256 (416) &                                                        253 (411) \\
  \hline
          ES1 &            $\sim$40ks &             \swire &                        740 (162) &       345 (133) &         259 (99) &                              257 (98) &                                              250 (72) &                                               182 (50) &                                                         180 (49) \\
  \hline
          XMM &        $\sim$50-100ks &             \swire &                      2167 (1083) &       348 (306) &        244 (226) &                             242 (224) &                                             295 (251) &                                              198 (178) &                                                        198 (178) \\
              &         $\sim$10-50ks &                    &                        371 (892) &       272 (902) &        214 (717) &                             212 (709) &                                             151 (409) &                                              111 (296) &                                                        111 (296)   
  \enddata
  \tablenotetext{a}{For distributions where we require X-ray detections, we use the X-ray completeness weights.}
\end{deluxetable*}

\begin{deluxetable*}{lrrrrrrr}  
  \tablecolumns{8}
  \tablewidth{0pt}
  \tablecaption{Overlap Fraction of Stern IR-AGNs and X-Ray AGNs with X-Ray Weights. \label{table:sternoverlap}}
  \tablehead{Field & \colhead{X-Ray Depth} & \colhead{IR Depth} & \colhead{\begin{sideways}\makecell[l]{Percent of IR-AGNs\\ that are X-ray-Detected}\end{sideways}} & \colhead{\begin{sideways}\makecell[l]{Percent of X-Ray\\ that are IR-AGN-Selected}\end{sideways}} & \colhead{\begin{sideways}\makecell[l]{Fraction of Total Sample\\detected by both}\end{sideways}} & \colhead{\begin{sideways}\makecell[l]{IR-AGNs Percent Increase\\to Total Sample Size}\end{sideways}} &    \colhead{\begin{sideways}\makecell[l]{X-Ray AGNs Percent Increase\\to Total Sample Size}\end{sideways}}  }
  \startdata
              CDFS &             $\sim$2Ms &             \goods &                                                                                  71.4\%$\pm$0.9\% &                                                                                  35.0\%$\pm$0.7\% &                                                                                 30.7\%$\pm$0.6\% &                                                                                    12.3\%$\pm$0.4\% &                                                                                      57.0\%$\pm$0.7\% \\
                   &                       &            \cosmos &                                                                                  87.7\%$\pm$1.0\% &                                                                                  20.3\%$\pm$0.6\% &                                                                                 19.7\%$\pm$0.6\% &                                                                                     2.8\%$\pm$0.2\% &                                                                                      77.5\%$\pm$0.6\% \\
                   &                       &             \swire &                                                                                      90\%$\pm$2\% &                                                                                   6.2\%$\pm$0.3\% &                                                                                  6.1\%$\pm$0.3\% &                                                                                     0.7\%$\pm$0.1\% &                                                                                      93.2\%$\pm$0.4\% \\
                   &           $\sim$200ks &             \goods &                                                                                      47\%$\pm$1\% &                                                                                      34\%$\pm$1\% &                                                                                 24.7\%$\pm$0.8\% &                                                                                    27.7\%$\pm$0.8\% &                                                                                      47.6\%$\pm$0.9\% \\
                   &                       &            \cosmos &                                                                                      82\%$\pm$1\% &                                                                                      38\%$\pm$1\% &                                                                                     35\%$\pm$1\% &                                                                                     7.7\%$\pm$0.6\% &                                                                                          57\%$\pm$1\% \\
                   &                       &             \swire &                                                                                      91\%$\pm$2\% &                                                                                  15.3\%$\pm$0.8\% &                                                                                 15.0\%$\pm$0.8\% &                                                                                     1.5\%$\pm$0.3\% &                                                                                      83.4\%$\pm$0.8\% \\
  \hline
            COSMOS &           $\sim$160ks &            \cosmos &                                                                                      74\%$\pm$1\% &                                                                                      35\%$\pm$1\% &                                                                                 30.9\%$\pm$1.0\% &                                                                                    11.0\%$\pm$0.6\% &                                                                                          58\%$\pm$1\% \\
                   &            $\sim$40ks &                    &                                                                                      57\%$\pm$2\% &                                                                                      53\%$\pm$2\% &                                                                                     38\%$\pm$1\% &                                                                                        29\%$\pm$1\% &                                                                                          33\%$\pm$1\% \\
  \hline
               ES1 &            $\sim$40ks &             \swire &                                                                                      72\%$\pm$2\% &                                                                                      34\%$\pm$2\% &                                                                                     30\%$\pm$2\% &                                                                                        11\%$\pm$1\% &                                                                                          59\%$\pm$2\% \\
  \hline
               XMM &        $\sim$50-100ks &             \swire &                                                                                      85\%$\pm$2\% &                                                                                  13.6\%$\pm$0.7\% &                                                                                 13.3\%$\pm$0.7\% &                                                                                     2.4\%$\pm$0.3\% &                                                                                      84.3\%$\pm$0.8\% \\
                   &         $\sim$10-50ks &                    &                                                                                      55\%$\pm$3\% &                                                                                      41\%$\pm$3\% &                                                                                     31\%$\pm$2\% &                                                                                        25\%$\pm$2\% &                                                                                          45\%$\pm$2\%   
  \enddata
\end{deluxetable*}

\begin{deluxetable*}{lrrrrrrr}  
  \tablecolumns{8}
  \tablewidth{0pt}
  \tablecaption{Overlap Fraction of Donley IR-AGNs and X-Ray AGNs with X-Ray Weights. \label{table:donleyoverlap}}
  \tablehead{Field & \colhead{X-Ray Depth} & \colhead{IR Depth} & \colhead{\begin{sideways}\makecell[l]{Percent of IR-AGNs\\ that are X-ray-Detected}\end{sideways}} & \colhead{\begin{sideways}\makecell[l]{Percent of X-Ray\\ that are IR-AGN-Selected}\end{sideways}} & \colhead{\begin{sideways}\makecell[l]{Fraction of Total Sample\\detected by both}\end{sideways}} & \colhead{\begin{sideways}\makecell[l]{IR-AGNs Percent Increase\\to Total Sample Size}\end{sideways}} &    \colhead{\begin{sideways}\makecell[l]{X-Ray AGNs Percent Increase\\to Total Sample Size}\end{sideways}}  }
  \startdata
              CDFS &             $\sim$2Ms &             \goods &                                                                                      83\%$\pm$1\% &                                                                                  18.4\%$\pm$0.5\% &                                                                                 17.7\%$\pm$0.5\% &                                                                                     3.6\%$\pm$0.3\% &                                                                                      78.7\%$\pm$0.6\% \\
                   &                       &            \cosmos &                                                                                      88\%$\pm$1\% &                                                                                  12.8\%$\pm$0.5\% &                                                                                 12.6\%$\pm$0.5\% &                                                                                     1.6\%$\pm$0.2\% &                                                                                      85.8\%$\pm$0.5\% \\
                   &                       &             \swire &                                                                                      91\%$\pm$2\% &                                                                                   5.0\%$\pm$0.3\% &                                                                                  4.9\%$\pm$0.3\% &                                                                                     0.5\%$\pm$0.1\% &                                                                                      94.6\%$\pm$0.3\% \\
                   &           $\sim$200ks &             \goods &                                                                                      60\%$\pm$2\% &                                                                                  19.8\%$\pm$0.8\% &                                                                                 17.5\%$\pm$0.8\% &                                                                                    11.7\%$\pm$0.6\% &                                                                                      70.8\%$\pm$0.9\% \\
                   &                       &            \cosmos &                                                                                      84\%$\pm$2\% &                                                                                  21.0\%$\pm$0.9\% &                                                                                 20.2\%$\pm$0.9\% &                                                                                     3.9\%$\pm$0.4\% &                                                                                      75.9\%$\pm$0.9\% \\
                   &                       &             \swire &                                                                                      89\%$\pm$2\% &                                                                                  10.5\%$\pm$0.7\% &                                                                                 10.4\%$\pm$0.7\% &                                                                                     1.2\%$\pm$0.2\% &                                                                                      88.4\%$\pm$0.7\% \\
  \hline
            COSMOS &           $\sim$160ks &            \cosmos &                                                                                      79\%$\pm$2\% &                                                                                  21.2\%$\pm$0.9\% &                                                                                 20.0\%$\pm$0.9\% &                                                                                     5.4\%$\pm$0.5\% &                                                                                      74.5\%$\pm$0.9\% \\
                   &            $\sim$40ks &                    &                                                                                      63\%$\pm$2\% &                                                                                      34\%$\pm$2\% &                                                                                     28\%$\pm$2\% &                                                                                        17\%$\pm$1\% &                                                                                          55\%$\pm$2\% \\
  \hline
               ES1 &            $\sim$40ks &             \swire &                                                                                      70\%$\pm$3\% &                                                                                      25\%$\pm$2\% &                                                                                     22\%$\pm$1\% &                                                                                         9\%$\pm$1\% &                                                                                          68\%$\pm$2\% \\
  \hline
               XMM &        $\sim$50-100ks &             \swire &                                                                                      81\%$\pm$3\% &                                                                                   9.1\%$\pm$0.6\% &                                                                                  8.9\%$\pm$0.6\% &                                                                                     2.1\%$\pm$0.3\% &                                                                                      89.0\%$\pm$0.7\% \\
                   &         $\sim$10-50ks &                    &                                                                                      52\%$\pm$3\% &                                                                                      30\%$\pm$2\% &                                                                                     23\%$\pm$2\% &                                                                                        22\%$\pm$2\% &                                                                                          55\%$\pm$2\%   
  \enddata
\end{deluxetable*}

\begin{deluxetable*}{lrrrrrrr}  
  \tablecolumns{8}
  \tablewidth{0pt}
  \tablecaption{Overlap Fraction of Stern IR-AGNs and X-Ray AGNs without X-Ray Weights. \label{table:sternoverlapnoweight}}
  \tablehead{Field & \colhead{X-Ray Depth} & \colhead{IR Depth} & \colhead{\begin{sideways}\makecell[l]{Percent of IR-AGNs\\ that are X-ray-Detected}\end{sideways}} & \colhead{\begin{sideways}\makecell[l]{Percent of X-Ray\\ that are IR-AGN-Selected}\end{sideways}} & \colhead{\begin{sideways}\makecell[l]{Fraction of Total Sample\\detected by both}\end{sideways}} & \colhead{\begin{sideways}\makecell[l]{IR-AGNs Percent Increase\\to Total Sample Size}\end{sideways}} &    \colhead{\begin{sideways}\makecell[l]{X-Ray AGNs Percent Increase\\to Total Sample Size}\end{sideways}}  }
  \startdata
              CDFS &             $\sim$2Ms &             \goods &                                                                                      46\%$\pm$1\% &                                                                                  19.9\%$\pm$0.7\% &                                                                                 16.1\%$\pm$0.6\% &                                                                                    18.9\%$\pm$0.6\% &                                                                                      65.0\%$\pm$0.8\% \\
                   &                       &            \cosmos &                                                                                      67\%$\pm$2\% &                                                                                  14.3\%$\pm$0.8\% &                                                                                 13.4\%$\pm$0.7\% &                                                                                     6.5\%$\pm$0.5\% &                                                                                      80.2\%$\pm$0.9\% \\
                   &                       &             \swire &                                                                                      79\%$\pm$3\% &                                                                                   6.1\%$\pm$0.5\% &                                                                                  6.0\%$\pm$0.5\% &                                                                                     1.6\%$\pm$0.3\% &                                                                                      92.4\%$\pm$0.6\% \\
                   &           $\sim$200ks &             \goods &                                                                                      36\%$\pm$1\% &                                                                                      33\%$\pm$1\% &                                                                                 20.8\%$\pm$0.9\% &                                                                                        38\%$\pm$1\% &                                                                                          42\%$\pm$1\% \\
                   &                       &            \cosmos &                                                                                      64\%$\pm$2\% &                                                                                      22\%$\pm$1\% &                                                                                     20\%$\pm$1\% &                                                                                    11.3\%$\pm$0.8\% &                                                                                          69\%$\pm$1\% \\
                   &                       &             \swire &                                                                                      81\%$\pm$3\% &                                                                                  10.0\%$\pm$0.8\% &                                                                                  9.7\%$\pm$0.8\% &                                                                                     2.3\%$\pm$0.4\% &                                                                                      87.9\%$\pm$0.9\% \\
  \hline
            COSMOS &           $\sim$160ks &            \cosmos &                                                                                      54\%$\pm$2\% &                                                                                      25\%$\pm$1\% &                                                                                     21\%$\pm$1\% &                                                                                        18\%$\pm$1\% &                                                                                          61\%$\pm$1\% \\
                   &            $\sim$40ks &                    &                                                                                      40\%$\pm$2\% &                                                                                      40\%$\pm$2\% &                                                                                     25\%$\pm$2\% &                                                                                        37\%$\pm$2\% &                                                                                          38\%$\pm$2\% \\
  \hline
               ES1 &            $\sim$40ks &             \swire &                                                                                      40\%$\pm$4\% &                                                                                      25\%$\pm$3\% &                                                                                     18\%$\pm$2\% &                                                                                        27\%$\pm$2\% &                                                                                          54\%$\pm$3\% \\
  \hline
               XMM &        $\sim$50-100ks &             \swire &                                                                                      71\%$\pm$3\% &                                                                                      13\%$\pm$1\% &                                                                                 12.0\%$\pm$1.0\% &                                                                                     4.8\%$\pm$0.6\% &                                                                                          83\%$\pm$1\% \\
                   &         $\sim$10-50ks &                    &                                                                                      37\%$\pm$3\% &                                                                                      33\%$\pm$3\% &                                                                                     21\%$\pm$2\% &                                                                                        36\%$\pm$3\% &                                                                                          43\%$\pm$3\%   
  \enddata
\end{deluxetable*}

\begin{deluxetable*}{lrrrrrrr}  
  \tablecolumns{8}
  \tablewidth{0pt}
  \tablecaption{Overlap Fraction of Donley IR-AGNs and X-Ray AGNs without X-Ray Weights. \label{table:donleyoverlapnoweight}}
  \tablehead{Field & \colhead{X-Ray Depth} & \colhead{IR Depth} & \colhead{\begin{sideways}\makecell[l]{Percent of IR-AGNs\\ that are X-ray-Detected}\end{sideways}} & \colhead{\begin{sideways}\makecell[l]{Percent of X-Ray\\ that are IR-AGN-Selected}\end{sideways}} & \colhead{\begin{sideways}\makecell[l]{Fraction of Total Sample\\detected by both}\end{sideways}} & \colhead{\begin{sideways}\makecell[l]{IR-AGNs Percent Increase\\to Total Sample Size}\end{sideways}} &    \colhead{\begin{sideways}\makecell[l]{X-Ray AGNs Percent Increase\\to Total Sample Size}\end{sideways}}  }
  \startdata
              CDFS &             $\sim$2Ms &             \goods &                                                                                      68\%$\pm$2\% &                                                                                  13.1\%$\pm$0.6\% &                                                                                 12.3\%$\pm$0.6\% &                                                                                     5.9\%$\pm$0.4\% &                                                                                      81.8\%$\pm$0.7\% \\
                   &                       &            \cosmos &                                                                                      70\%$\pm$3\% &                                                                                   9.4\%$\pm$0.7\% &                                                                                  9.0\%$\pm$0.6\% &                                                                                     3.9\%$\pm$0.4\% &                                                                                      87.1\%$\pm$0.7\% \\
                   &                       &             \swire &                                                                                      82\%$\pm$3\% &                                                                                   5.7\%$\pm$0.5\% &                                                                                  5.6\%$\pm$0.5\% &                                                                                     1.2\%$\pm$0.2\% &                                                                                      93.1\%$\pm$0.6\% \\
                   &           $\sim$200ks &             \goods &                                                                                      50\%$\pm$2\% &                                                                                      21\%$\pm$1\% &                                                                                 17.3\%$\pm$0.9\% &                                                                                    17.3\%$\pm$0.9\% &                                                                                          65\%$\pm$1\% \\
                   &                       &            \cosmos &                                                                                      72\%$\pm$3\% &                                                                                  15.8\%$\pm$1.0\% &                                                                                 14.9\%$\pm$0.9\% &                                                                                     5.9\%$\pm$0.6\% &                                                                                          79\%$\pm$1\% \\
                   &                       &             \swire &                                                                                      83\%$\pm$3\% &                                                                                   9.0\%$\pm$0.8\% &                                                                                  8.9\%$\pm$0.8\% &                                                                                     1.9\%$\pm$0.4\% &                                                                                      89.3\%$\pm$0.8\% \\
  \hline
            COSMOS &           $\sim$160ks &            \cosmos &                                                                                      66\%$\pm$3\% &                                                                                      19\%$\pm$1\% &                                                                                     17\%$\pm$1\% &                                                                                     9.2\%$\pm$0.8\% &                                                                                          73\%$\pm$1\% \\
                   &            $\sim$40ks &                    &                                                                                      51\%$\pm$3\% &                                                                                      30\%$\pm$2\% &                                                                                     23\%$\pm$2\% &                                                                                        23\%$\pm$2\% &                                                                                          54\%$\pm$2\% \\
  \hline
               ES1 &            $\sim$40ks &             \swire &                                                                                      38\%$\pm$4\% &                                                                                      19\%$\pm$2\% &                                                                                     14\%$\pm$2\% &                                                                                        23\%$\pm$2\% &                                                                                          63\%$\pm$3\% \\
  \hline
               XMM &        $\sim$50-100ks &             \swire &                                                                                      71\%$\pm$4\% &                                                                                  11.1\%$\pm$1.0\% &                                                                                 10.6\%$\pm$0.9\% &                                                                                     4.2\%$\pm$0.6\% &                                                                                          85\%$\pm$1\% \\
                   &         $\sim$10-50ks &                    &                                                                                      35\%$\pm$4\% &                                                                                      26\%$\pm$3\% &                                                                                     17\%$\pm$2\% &                                                                                        32\%$\pm$3\% &                                                                                          50\%$\pm$3\%   
  \enddata
\end{deluxetable*}

\end{document}